\documentclass{article}
\usepackage{arxiv}
\usepackage{dcolumn}
\usepackage{mathptmx} 
\usepackage{multirow}
\usepackage{} 

\usepackage[utf8]{inputenc} 
\usepackage[T1]{fontenc}    
\usepackage{hyperref}       
\usepackage{url}            
\usepackage{booktabs}       
\usepackage{amsfonts}       
\usepackage{nicefrac}       
\usepackage{microtype}      
\usepackage{lipsum}
\usepackage{amssymb,amsmath,epsfig}
\newcommand{\beq}{\begin{equation}}
\newcommand{\eeq}{\end{equation}}
\newcommand{\bea}{\begin{eqnarray}}
\newcommand{\eea}{\end{eqnarray}}

\begin{document}
\title{Shadow cast and center of mass energy in a charged Gauss-Bonnet-AdS black hole}
\author{{Muhammad Zahid\thanks{zahid.m0011@gmail.com}}\,, {Saeed Ullah Khan\thanks{saeedkhan.u@gmail.com}}\,, and {Jingli Ren\thanks{Corresponding author: renjl@zzu.edu.cn}}
\vspace{0.2cm} \\\vspace{0.1cm}
Henan Academy of Big Data/School of Mathematics and Statistics, Zhengzhou University Zhengzhou\\ 450001,  China.}
\date{}
\maketitle
\begin{abstract}
This work is devoted to the exploration of shadow cast and center of mass energy in the background of a 4-dimensional charged Gauss-Bonnet AdS black hole. On investigating particle dynamics, we have examined BH's metric function. Whereas, with the help of null geodesics, we pursue to calculate the celestial coordinates and the shadow radius of the black hole. We have made use of the hawking temperature to study the energy emission rate. Moreover, we have explored the center of mass energy and discussed its characteristics under the influence of spacetime parameters. For a better understanding, we graphically represent all of our main findings. The acquired result shows that both charge and AdS radius ($l$) decrease the shadow radius, while the Gauss-Bonnet coupling parameter $\alpha$ increases the shadow radius in AdS spacetime. On the other hand, both $Q$ and $\alpha$ result in diminishing the shadow radius in asymptotically flat spacetime. Finally, we investigate the energy emission rate and center of mass energy under the influence of $Q$ and $\alpha$.\\
\end{abstract}
\section{Introduction}
\label{sec:1}
After the discovery of Einstein's General relativity theory \cite{a1}, black holes (BHs) investigations caught a considerable interest of many researchers \cite{r1,r2,Stuchlikschee,r3,r4}. This theory explains how gravitational field affects the relation of time and space, as well as how quickly could an observer observed change in an object. More precisely, we can conclude that the gravitational field created by the existence of matter bends spacetime, which manages the motion of objects in space. To tackle fundamental problems such as quantum gravity and singularity problems, different types of modified gravity theories have been discovered in the past decades. Among them, the most promising approach is the Gauss-Bonnet (GB) gravity with higher curvature correction. Since the fact is well-known that for d-dimensional spacetime with d$>$4, there exist different types of static and spherically symmetric BH solutions. But in the 4-dimensional case, the GB term is a total derivative, therefore, will not contribute to the gravitational dynamics. For 4-dimensional spacetime, the generally modified gravity in which graviton propagation takes place only for the massless particle has been carried out recently by Glavan and Lin \cite{a2}. In this case, the GB term has a non-trivial contribution to the gravitational dynamics. This led to the discovery of non-trivial four-dimensional static and spherically symmetric solutions of BHs. R.A Kenoplya and A.F Zinhailo \cite{a3} carried out the study of Quasi-normal modes in which the variation of GB parameter shows more sensitivity for the damping rate than that of real one.

Gravitational lensing is a phenomenon in which an object moving around a BH having critical radius falls into it. But if the moving object is a photon, then it forms shadow in the plane which can be observed at infinity and was first studied by Bardeen \cite{a4}. The circular shadow can be obtained if the BH is spherically symmetric \cite{a5,a6}, whereas a deformed shadow will be cast in the case of rotating BHs \cite{a7,a8}. For Schwarzschild BH, at first, the angular radius was studied by \cite{a2}, while Hioki et. al \cite{a9} later introduced the two observable parameters. Due to which the investigation of BH shadows triggered a wide variety of cases \cite{a10,ScheeStucklik,a11,a12,a13,a14,a15,a16,a17,a18,a19,a20}. 

Recently, Khan and Ren \cite{r5} by studying a spinning charged BH in quintessential dark energy found that the quintessence parameter contributes to the shadow radius whereas, diminishing the distortion effects in the shadow. The exploration of rotating Randall-Sundrum braneworld BH with a cosmological constant demonstrate that the cosmological constant increases, while the tidal charge decrease and distort shadow of the BH \cite{Khan5}. In the case of spherically symmetric BH, shadow cast reflect a close relationship between the coupling 
parameter and size of the shadow \cite{a3}. Moreover, the radii of an innermost stable circular orbit, photon sphere, and 
BH horizon represent a decreasing function of the GB parameter \cite{a24}. On considering higher dimensions, Pas et al. \cite{a26} studied shadow cast in the GB gravity spacetime for a non-rotating BH, and \cite{a27,a28} studied it by considering the spinning BH. Besides, the Event Horizon Telescope (EHT) is an ongoing project working on gathering data from the Milky way, as well as from the nearby galaxies of the supermassive BHs \cite{a21}. In 2019, a group of researchers from EHT published the image of M87 \cite{a22}. This discovery enabled everyone to see BH's image for the first time. 
To predict the possible observable the shadow of warm holes was investigated in \cite{a23}.

Center of mass energy (CME) is the process of the formation of new particles due to energy formed by the colliding particles in the center of mass frame. CME usually depends on factors such as particle's nature, astrophysical objects, and on the gravitational field around the object. In 2009, Banados, Silk and West \cite{a29} investigated the influence of infinite energy growth in the center of mass frame. This phenomenon is called the BSW effect. Firstly, the BSW effect was studied in the context of an extreme Kerr BH. The particles collide near the horizon of a BH are blue-shifted because of infinite energy. For the non-extremal case, Lake \cite{a30} obtained a finite CME near the Kerr BH horizon. However, when the collision of two particles of equal masses takes place near horizons, then one can obtain infinite CME in the non-extremal case as well \cite{a31}. In the last decade, researchers have explored the BSW effect for different types of rotating and non-rotating BHs \cite{a32,a33,a34,a35}.

The layout of this work can be followed as, in section (\ref{sec:2}), we will drive geodesic equations in the background of GB-AdS BH. After that, BH's shadow and energy emission rate for both GB-AdS and asymptotically flat spacetime will be studied in section (\ref{sec:3}). However, section (\ref{sec:4}) will deal with the investigation of CME in GB-AdS BH. In the last section (\ref{sec:5}), we will conclude our findings and observations.
\section{Gauss–Bonnet-AdS black hole}
\label{sec:2}
In GB gravity, generally, there are two types of BH solutions for $d\geq5$ namely, the static and spherically symmetric ones \cite{a36,a37,a38,a39,a40}. Whereas in four-dimensional spacetime, the GB term is a total derivative, thus has no contribution to the gravitational dynamics. Glavan and Lin \cite{a2} by considering $d\rightarrow4$ and re-scaling the GB coupling parameter $\alpha\rightarrow\frac{\alpha}{d-4}$ found the non-trivial four-dimensional BH solution.

\begin{figure*}
\begin{minipage}[b]{0.58\textwidth} \hspace{-0.cm}
\includegraphics[width=0.8\textwidth]{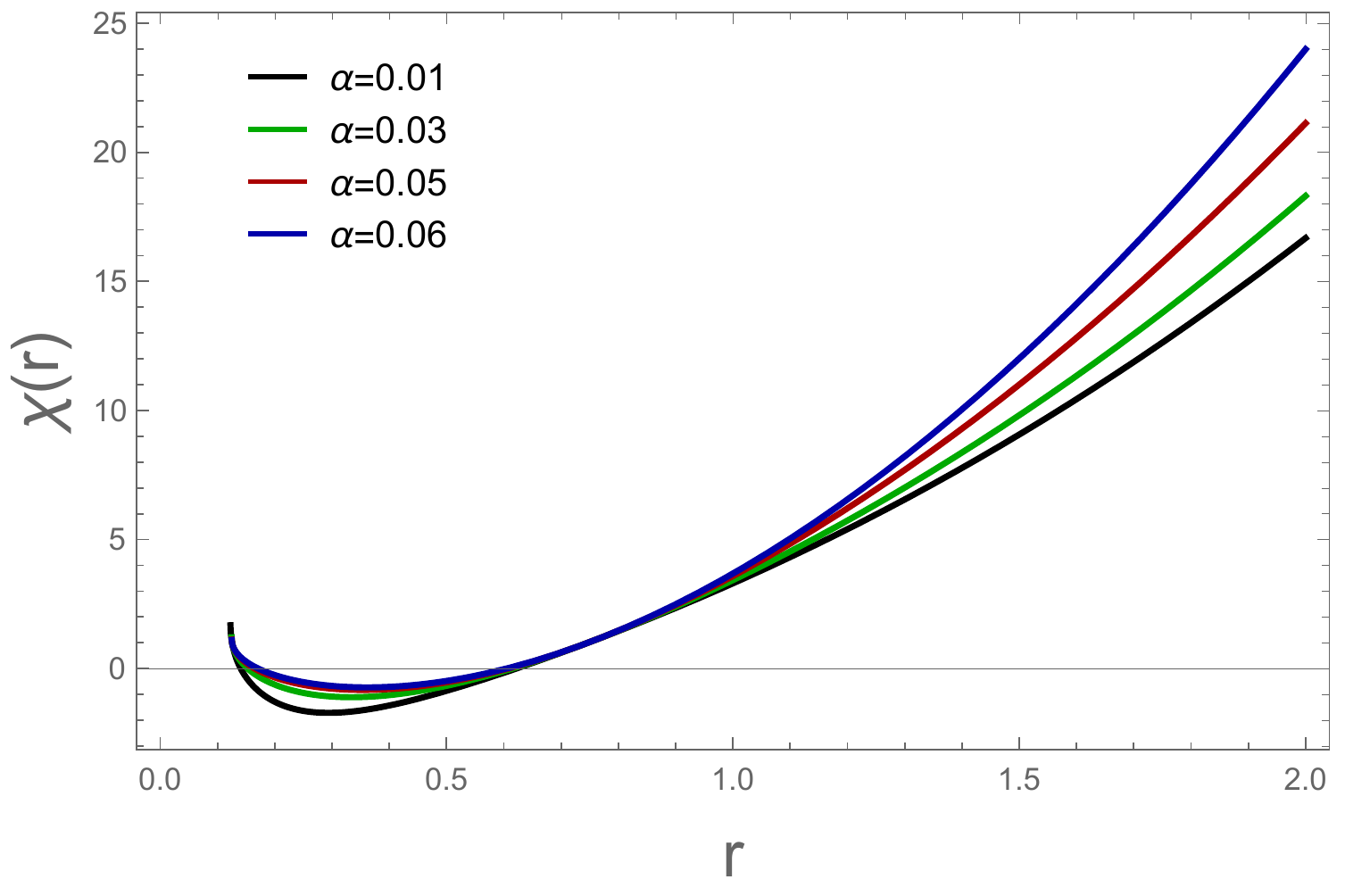}
\end{minipage}
\vspace{0cm}
\begin{minipage}[b]{0.58\textwidth} \hspace{-1.5cm}
\includegraphics[width=0.8\textwidth]{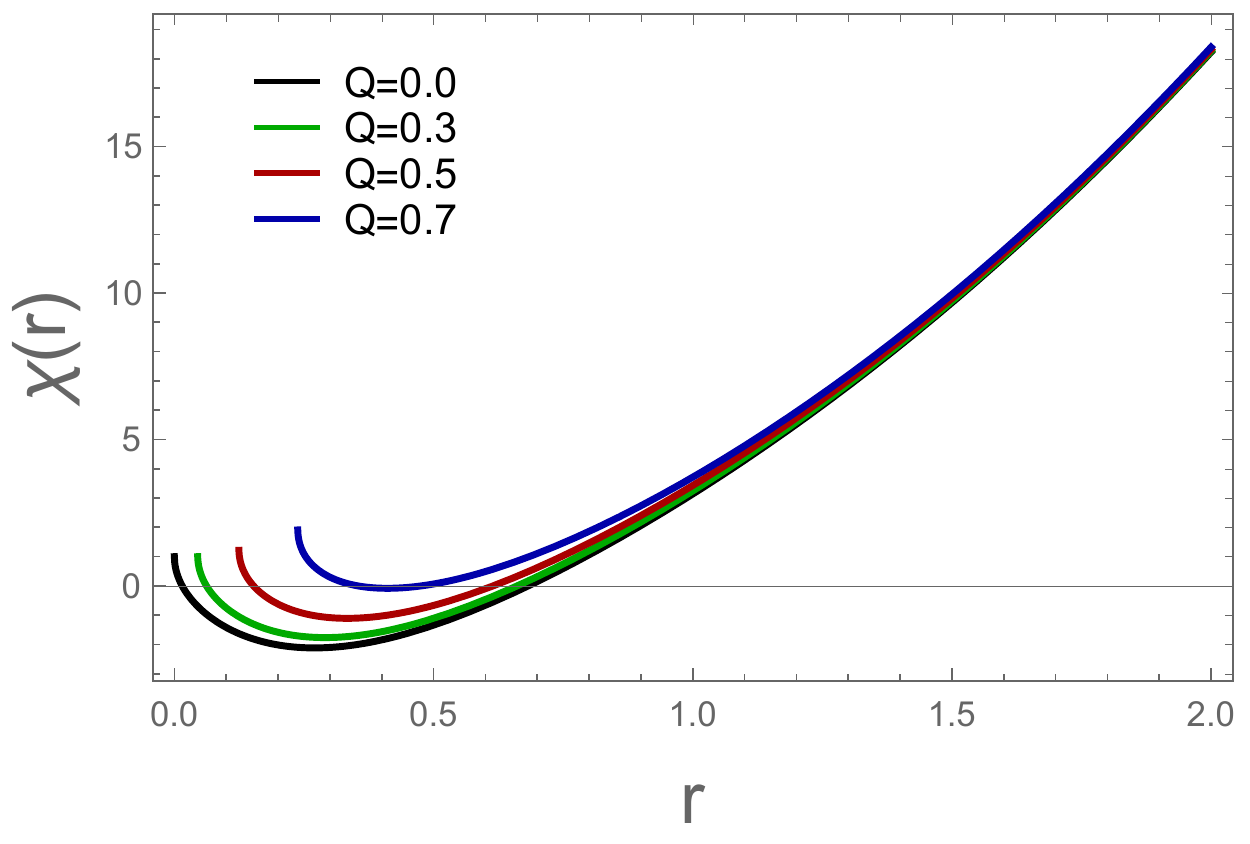}
\end{minipage}
\caption{Graphical representation of the metric function at $Q=0.5$ (left panel), while at $\alpha=0.03$ (right panel) of the GB-AdS BH ($l=0.5$).}\label{Horizons1}
\end{figure*}

\begin{figure*}
\begin{minipage}[b]{0.58\textwidth} \hspace{-0.cm}
\includegraphics[width=0.8\textwidth]{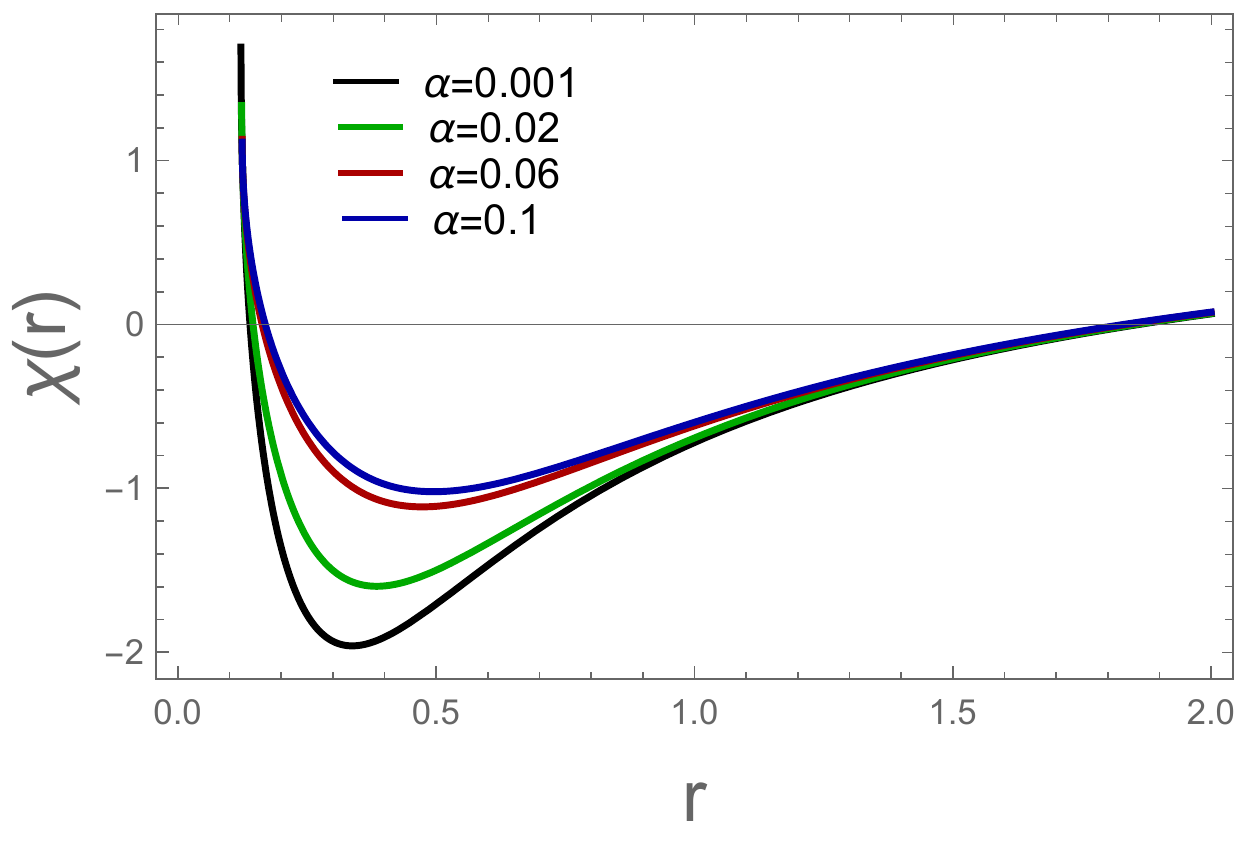}
\end{minipage}
\begin{minipage}[b]{0.58\textwidth} \hspace{-1.5cm}
\includegraphics[width=0.8\textwidth]{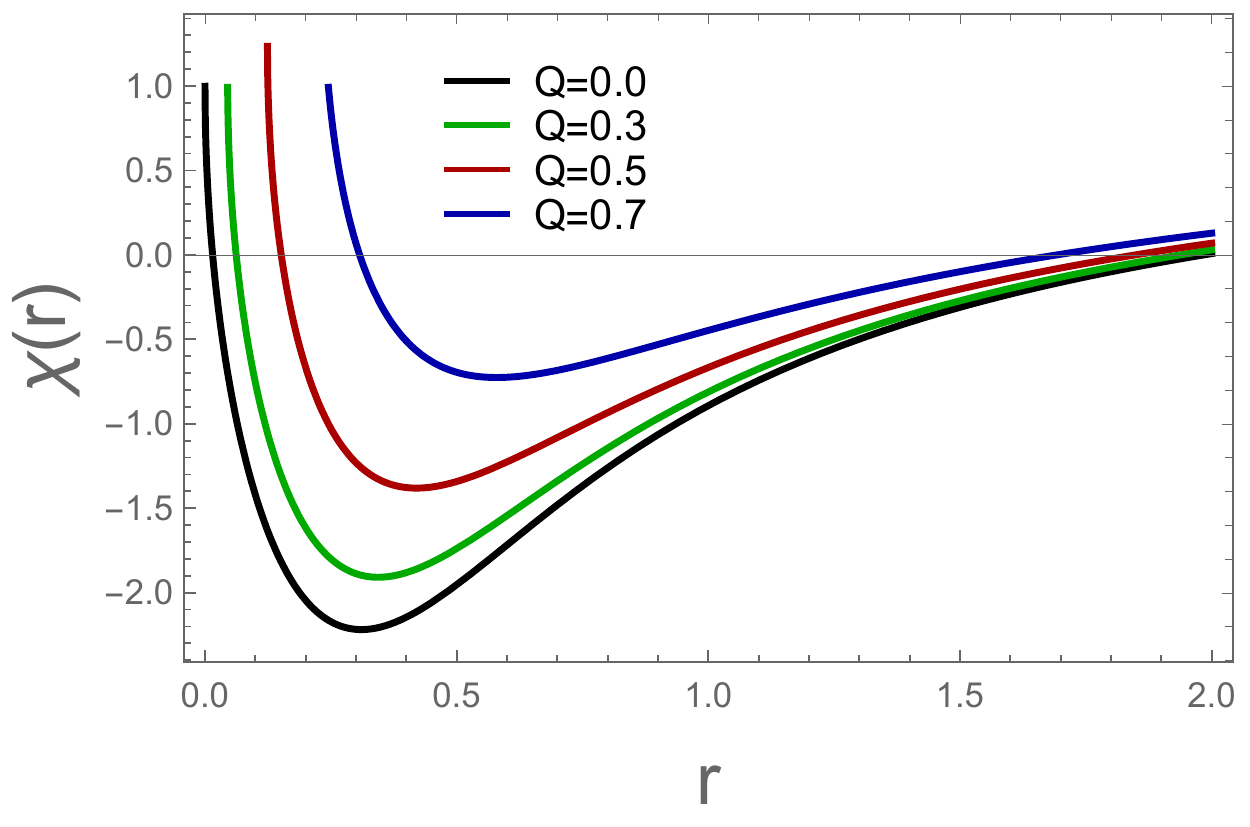}
\end{minipage}
\caption{Graphical representation of the metric function at $Q=0.5$ (left panel), while at $\alpha=0.03$ (right panel) of the Asymptotically flat spacetime BH ($l\rightarrow\infty$).}\label{Horizons2}
\end{figure*}

For the d-dimensional spacetime, the action of GB gravity can be expressed by
\begin{equation}\label{e1}
S=\frac{1}{16\pi}\int d^{d}x \left(R +\frac{(d-1)(d-2}{l^{2}}+\frac{\alpha}{d-4}L_{GB}-F_{\mu\nu}F^{\mu\nu}\right),
\end{equation}
with
\begin{eqnarray}\label{e2}
L_{GB}=R_{\mu\nu\rho\sigma}R^{\mu\nu\rho\sigma}-4R_{\mu\nu}R^{\mu\nu}+R^{2}.
\end{eqnarray}
Here $F_{\mu\nu}$ is a Maxwell tensor and $l$ represents AdS radius and can be related to the cosmological constant $\wedge$, as
\begin{equation}\label{e3}
\wedge=\frac{(d-1)(d-2)}{2l^{2}}.
\end{equation}
After adjusting limit $d\rightarrow4$, the field equation obtained from this action read as \cite{a25}
\begin{equation}\label{e4}
dS^{2}=-\chi(r)dt^{2}+\chi^{-1}(r) dr^{2}+r^{2}d\theta^{2}+r^{2}\sin^{2}\theta d\phi^{2},
\end{equation} 
with
\begin{equation}\label{e5}
\chi(r)=1+\frac{r^{2}}{2\alpha}\left(1-\sqrt{1+4\alpha\left(\frac{2M}{r^{3}}-\frac{Q^{2}}{r^{4}}-\frac{1}{l^{2}}\right)}\right).
\end{equation}
In Eq. \eqref{e5}, $Q$ and $M$ denote charge and mass parameters of the BH, respectively. By taking limit $\alpha\rightarrow0$, one can recover the Reissner-Nordst{\"o}rm-AdS BH. Similarly setting $l\rightarrow\infty$ and $Q\rightarrow0$, results in static and spherically symmetric BH solutions \cite{a2}. The behavior of metric function under the influence of different parameters is shown in Figs. \ref{Horizons1} and \ref{Horizons2}, respectively for the GB-AdS and asymptotically flat spacetime BHs. It is clearly shown that metric function increases with both of the coupling and charge parameters in case of GB-AdS spacetime BH.  While in the case of asymptotically flat spacetime BH the metric function shows the same behavior but with different shape of its horizons.
\subsection{Geodesic equations}
\label{2.1}
\begin{figure*}
\begin{minipage}[b]{0.58\textwidth} \hspace{-0.cm}
\includegraphics[width=0.8\textwidth]{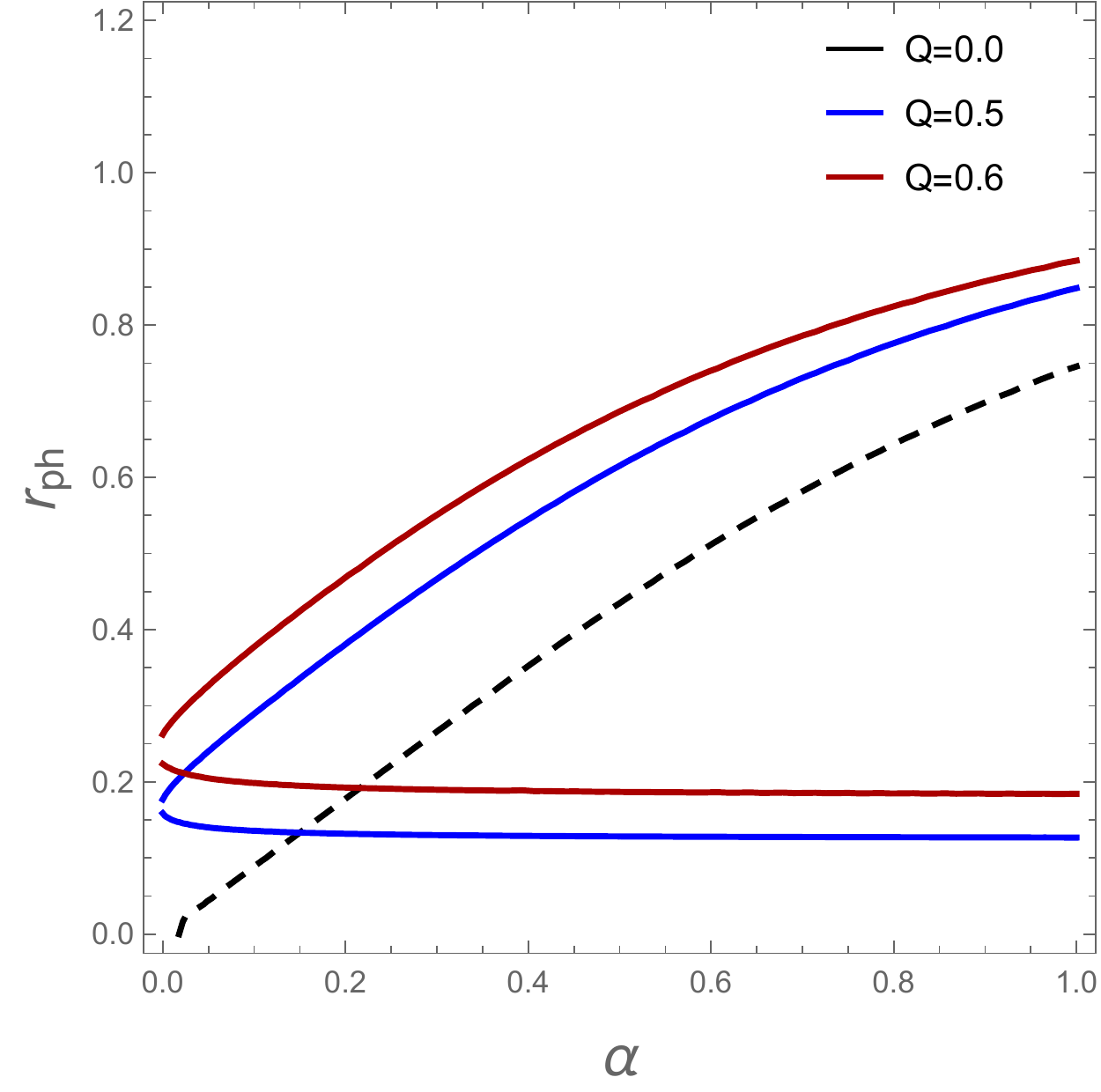}
\end{minipage}
\vspace{0cm}
\begin{minipage}[b]{0.58\textwidth} \hspace{-1.5cm}
\includegraphics[width=0.8\textwidth]{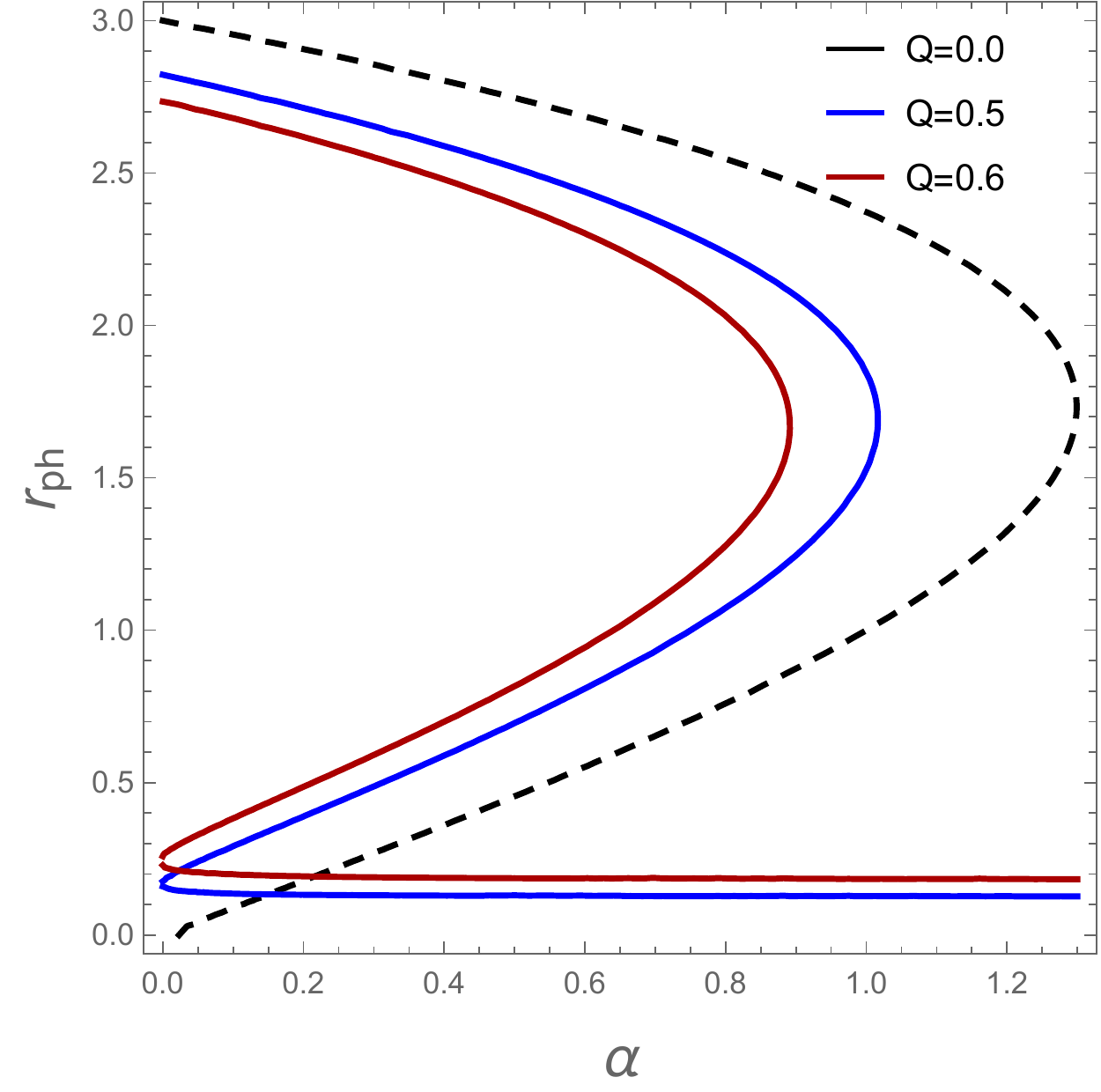}
\end{minipage}
\caption{Graphical representation of photon orbits along $\alpha$ at different values of $Q$ in the AdS (Left) and asymptotically flat spacetime (Right).}\label{Photon}
\end{figure*}

Geodesic motion around a BH is the main factor in predicting the possible structure of BH shadow. In this section, we will determine the geodesic equations around GB-AdS BH. For this purpose, let us assume a test particle moving around the GB-AdS BH with a rest mass of $m_{0}$. The problem of finding geodesic becomes simplified after determining the constant of motion associated with the symmetries direction. Let $k^{\nu}$ and $u^{\nu}=\frac{dx^{\nu}}{d\tau}$, be the vectors along symmetry direction and tangent vector along the curve $x^{\nu}=x^{\nu}(\tau)$, respectively here $\tau$ is an affine parameter. Now if the trajectory $x^{\nu}$ is a geodesic \cite{a41} then using the killing vectors, we have
\begin{equation}\label{e6}
k^{\nu}u_{\nu}=constant.
\end{equation}
Now for time independent metric coefficient there exists a time-like killing vector $k^{\nu}=(1,0,0,0)$. Then by using equation \eqref{e6}, we have
\begin{eqnarray}\label{e7}
k^{0}u^{0}=u^{0}=-\mathcal{E},
\end{eqnarray}
Similarly, for the $\phi$ direction, the killing vector can be written as $k^{\nu}=(0,0,0,1)$. Therefore, we have
\begin{eqnarray}\label{e8}
k^{3}u_{3}=u_{3}=L,
\end{eqnarray}
here $\mathcal{E}$ and $L$, represents the relativistic energy and angular momentum per unit mass of the particle, respectively. Now using Eqs. \eqref{e7} and \eqref{e8}, one can obtain the geodesic equation as follows
\begin{align}\label{e9}
\begin{split}
u^{0}=g^{0\mu}u_{\mu}=g^{00}u_{0}=\frac{\mathcal{E}}{\chi(r)},\\
u^{3}=g^{3\mu}u_{\mu}=g^{33}u_{3}=\frac{L}{r^{2} \sin^{2}\theta}.
\end{split}
\end{align}
Finally, Eq. \eqref{e9} can be written as
\begin{align}\label{e10}
\begin{split}
\frac{dt}{d\tau}=\frac{\mathcal{E}}{\chi(r)},\\
\frac{d\phi}{d\tau}=\frac{L}{r^{2} \sin^{2}\theta}.
\end{split}
\end{align}
In order to obtain the other two geodesic equations, we will adopt the Hamilton-Jacobi method. Therefore, we have
\begin{eqnarray}\label{e11}
\frac{\partial H}{\partial \tau}+\frac{1}{2}g^{\mu\sigma}\frac{\partial H}{\partial x^{\mu}}\frac{\partial H}{\partial x^{\sigma}}=0.
\end{eqnarray}
Assuming the ansatz of the form \cite{a42}, we have
\begin{equation}\label{e12}
H=H_{r}(r)-\mathcal{E} t+L\phi+H_{\theta}(\theta)+\frac{1}{2}m_{0}^{2}\tau.
\end{equation}
In Eq.\eqref{e12}, $H_{r}(r)$ and $H_{\theta}(\theta)$, represents the function of $r$ and $\theta$, respectively. Now by putting the Jacobi action \eqref{e11}, into the Hamilton-Jacobi equation \eqref{e12}, we have
\begin{align}\label{e13}
\begin{split}
H_{r}(r)=\int^{r}\frac{\sqrt{R(r)}}{r^{2}\chi(r)}dr,\\
H_{\theta}(\theta)=\int^{r}\sqrt{W(\theta})d\theta,
\end{split}
\end{align}
in which
\begin{eqnarray}\label{e14}
R(r)&=&r^{4}\mathcal{E}^{2}-r^{2}\left(r^{2}m_{0}^{2}+\kappa+L^{2}\right)\chi(r),\\
W(\theta)&=&\kappa-L^{2}\cot^{2}\theta.
\end{eqnarray}
Here $\kappa$ represents Carter constant. Now using the relations $S_{\theta}=\frac{\partial H}{\partial\theta}=\frac{\partial H_{\theta}}{\partial\theta}$ and $S_{r}=\frac{\partial H}{\partial r}=\frac{\partial H_{r}}{\partial r}$. Particle's geodesic equations of motion in the background of GB-AdS BH can be expressed as
\begin{equation}\label{e15}
r^{2}\frac{d\theta}{d\tau}=\sqrt{\kappa-L^{2}\cot^{2}\theta}=\sqrt{W(\theta)},
\end{equation}
\begin{equation}\label{e16}
r^{2}\frac{d r}{d\tau}=\sqrt{r^{4}\mathcal{E}^{2}-r^{2}\left(r^{2}m_{0}^{2}+\kappa+L^{2}\right)\chi(r)}=\sqrt{R(r)}.
\end{equation}
The above Eqs. \eqref{e15} and \eqref{e16}, are the required geodesic equations along the direction of $\theta$ and $r$, respectively. Moreover, by taking $m_{0}^{2}=0$ one can obtain the following geodesic equations traced by photon around the BH \cite{a26}.
\begin{equation}\label{e17}
r^{2}\frac{d\theta}{d\tau}=\sqrt{\kappa-L^{2}\cot^{2}\theta}=\sqrt{W(\theta)},
\end{equation}
\begin{equation}\label{e18}
r^{2}\frac{d r}{d\tau}=\sqrt{r^{4}\mathcal{E}^{2}-r^2\left(\kappa+L^{2}\right)\chi(r)}=\sqrt{R(r)}.
\end{equation}
Equation \eqref{e18} can be written in the following equivalent form as
\begin{equation}\label{e19}
\left(\frac{d r}{d \tau}\right)^{2}+V(r)=0,
\end{equation}
where $V(r)$, is the effective potential and is given by
\begin{equation}\label{e20}
V(r)=\frac{\chi(r)}{r^{2}}\left(\kappa+L^{2}\right)-\mathcal{E}^{2}.
\end{equation}
Next to find the unstable circular orbits we need to impose the following conditions
\begin{eqnarray}\label{e21}
V(r)\vert_{r=r_{ph}}=V^{'}(r)\vert_{r=r_{ph}}=0,
\end{eqnarray}
here $r=r_{ph}$, represent photon radius while $'$ denotes derivative with respect to $r$. Therefore, the condition $V(r_{ph})=0$ implies that
\begin{equation}\label{e22}
\frac{r_{ph}^{2}}{\chi(r_{ph})}=\zeta+\gamma^{2}.
\end{equation}
In Eq. \eqref{e22}, we used the Chandrasekhar constant definition \cite{a42}, i.e., $\zeta={\kappa}/{\mathcal{E}^{2}}$ and $\gamma={L}/{\mathcal{E}}$. Similarly $V^{'}(r_{ph})=0$, leads us to
\begin{eqnarray}\label{e24}
r_{ph}\chi^{'}(r_{ph})-2\chi(r_{ph})=0.
\end{eqnarray}
From Eq. \eqref{e5}, we have
\begin{eqnarray}\label{e25}
\chi^{'}(r_{ph})=\frac{4 r^{3}\alpha + l^{2}\left(-2 M\alpha + r^{3}\left(-1+\sqrt{1-\frac{4\alpha}{l^{2}}-\frac{4(Q^{2}-2 M r)\alpha}{r^{4}}}\right)\right)}{l^{2}r^{2}\alpha\sqrt{1-\frac{4\alpha}{l^{2}}-\frac{4(Q^{2}-2 M r)\alpha}{r^{4}}}}.
\end{eqnarray}
Thus, using Eqs. \eqref{e5} and \eqref{e25}, Eq. \eqref{e24} simplifies to
\begin{eqnarray}\label{e26}
\left(1-\frac{4\alpha}{l^{2}}\right)r_{ph}^{4}-9 r_{ph}^{2} M^{2}+4(2\alpha+3Q^{2})M r_{ph}-4(\alpha + Q^{2})Q^{2}=0.
\end{eqnarray}
Equation \eqref{e26}, can be solved analytically for $Q=0$, however, for $Q\neq0$, we solve it numerically. For this purpose, we need to fixed the values of $l$, $M$ and $\alpha$, then numerically solve it for different values of the Charge parameter $Q$ in order to get the values of $r_{ph}$. After setting limit $l\rightarrow\infty$, one can find the radius of photon in case of asymptotically flat spacetime. Therefore, in the case of asymptotically flat spacetime Eq. \eqref{e26}, can be written as
\begin{eqnarray}\label{e27}
r_{ph}^{4}-9 r_{ph}^{2} M^{2}+4(2\alpha+3Q^{2})M r_{ph}-4(\alpha + Q^{2})Q^{2}=0.
\end{eqnarray} 
Similarly, for chargeless case ($Q=0$), Eq. \eqref{e27} can be solved analytically but for $Q\neq0$, we need to solve it numerically. The graphical behavior of photon orbit in both spacetimes are shown in Fig. \ref{Photon} for different values of the charge parameter $Q$. The result shows that the radii of photon orbit increases by increasing the value of $Q$ in AdS spacetime. However, the BH charge diminishing the radii of photon orbit in the case of asymptotically flat spacetime.
\section{Black hole shadow}
\label{sec:3}
Consider a massless particle, such as a photon, emitted from an object moving towards the BH located between an observer and a bright object. The possible outcomes for the trajectories of the photons are; (a) scattered away; (b) falls into the BH; and (c) critical geodesic separating (a) and (b). In this case, the observer can only observe the scattering of a photon from the BH. However, a dark region will be formed in the case, when photon falls into the BH. This dark region is named as shadow of the BH.
In this section, our main goal is to explore shadow cast by a 4-dimensional charged GB-AdS BH spacetime. For this purpose, first of all, we need to define the celestial coordinates \cite{a20} as
\begin{align}\label{e28}
\begin{split}
\lambda=\lim_{r \to \infty}-\left(r^{2} \sin\theta\frac{d\phi}{dr}\right),\\
\xi=\lim_{r \to \infty}\left(r^{2}\frac{d\theta}{dr}\right).
\end{split}
\end{align}
In Eq. \eqref{e28}, $\lambda$ and $\xi$ denote the apparent perpendicular distance of shadow from the axis of symmetry and its projection on the equatorial plane, respectively. Whereas, $\theta$ represents the inclination angle between the axis of symmetry and the observer's line of sight, while $r_{0}$ is the distance between BH and observer. Now by using the geodesic equations, we can obtain the values of ${d\phi}/{dr}$ and ${d\theta}/{dr}$, as
\begin{figure*}
\begin{minipage}[b]{0.58\textwidth} \hspace{-0.cm}
\includegraphics[width=0.8\textwidth]{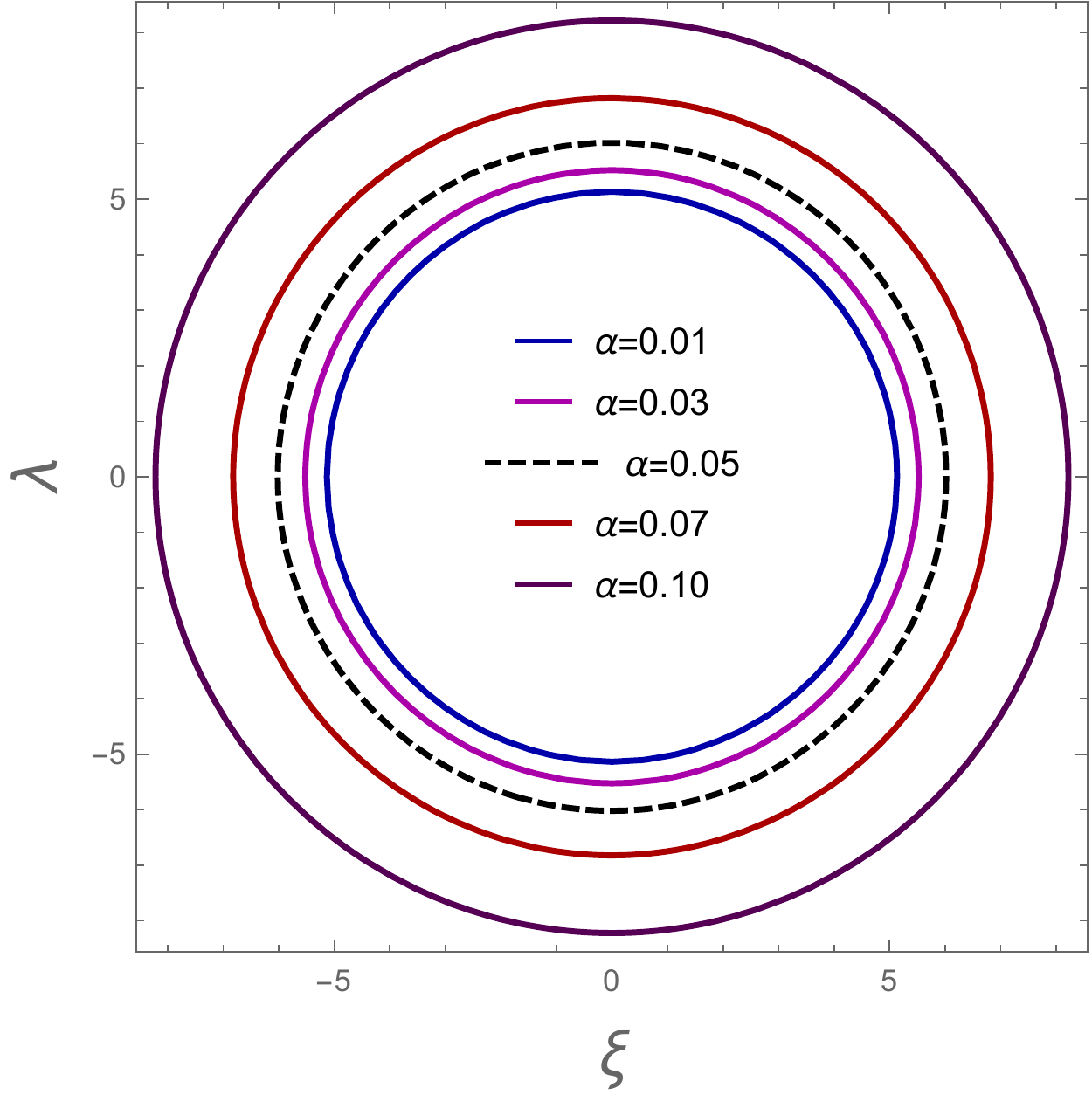}
\end{minipage}
\vspace{0.25cm}
\begin{minipage}[b]{0.58\textwidth} \hspace{-1.5cm}
\includegraphics[width=0.8\textwidth]{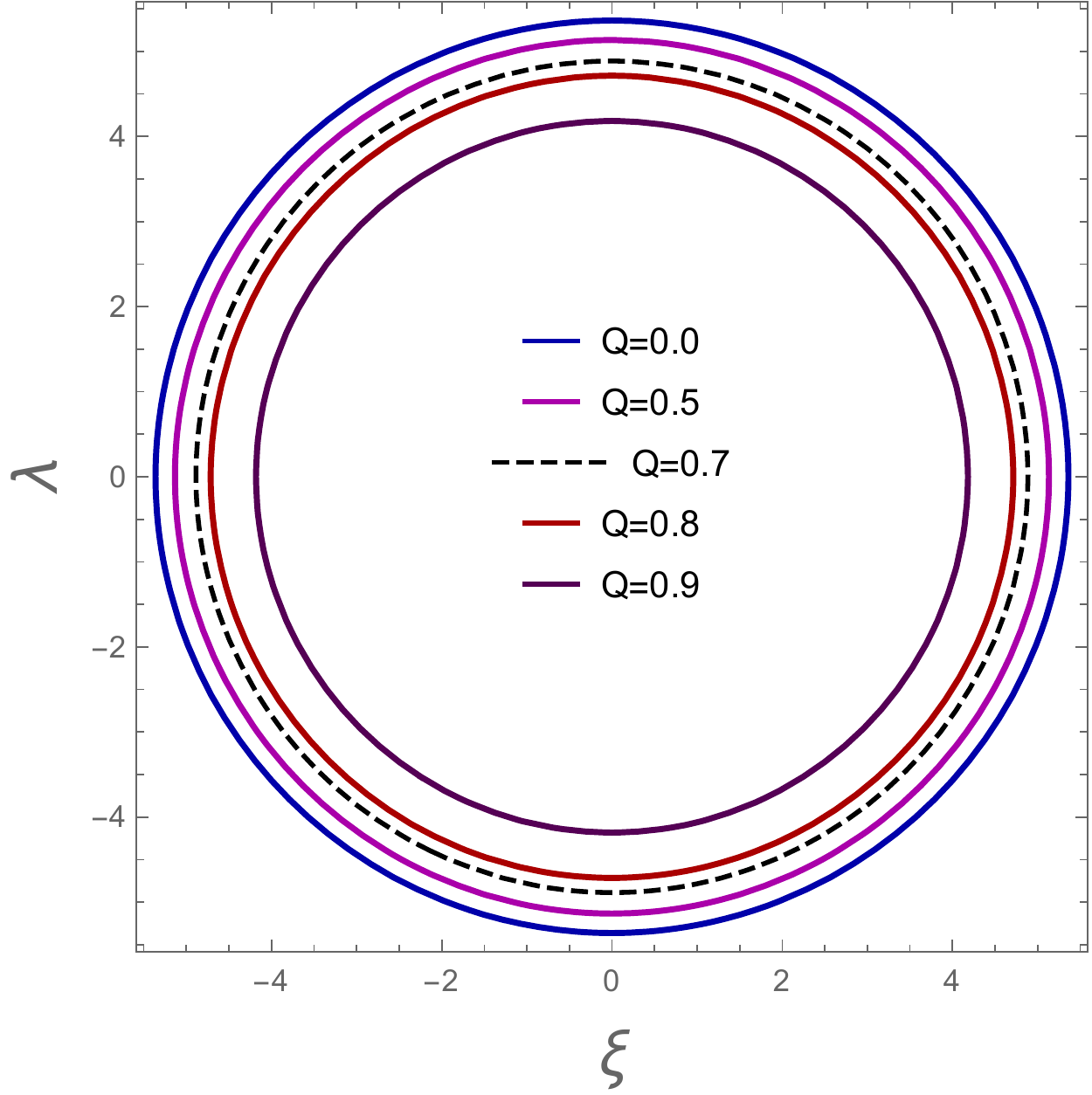}
\end{minipage}
\begin{minipage}[b]{0.58\textwidth} \hspace{-0.cm}
\includegraphics[width=0.8\textwidth]{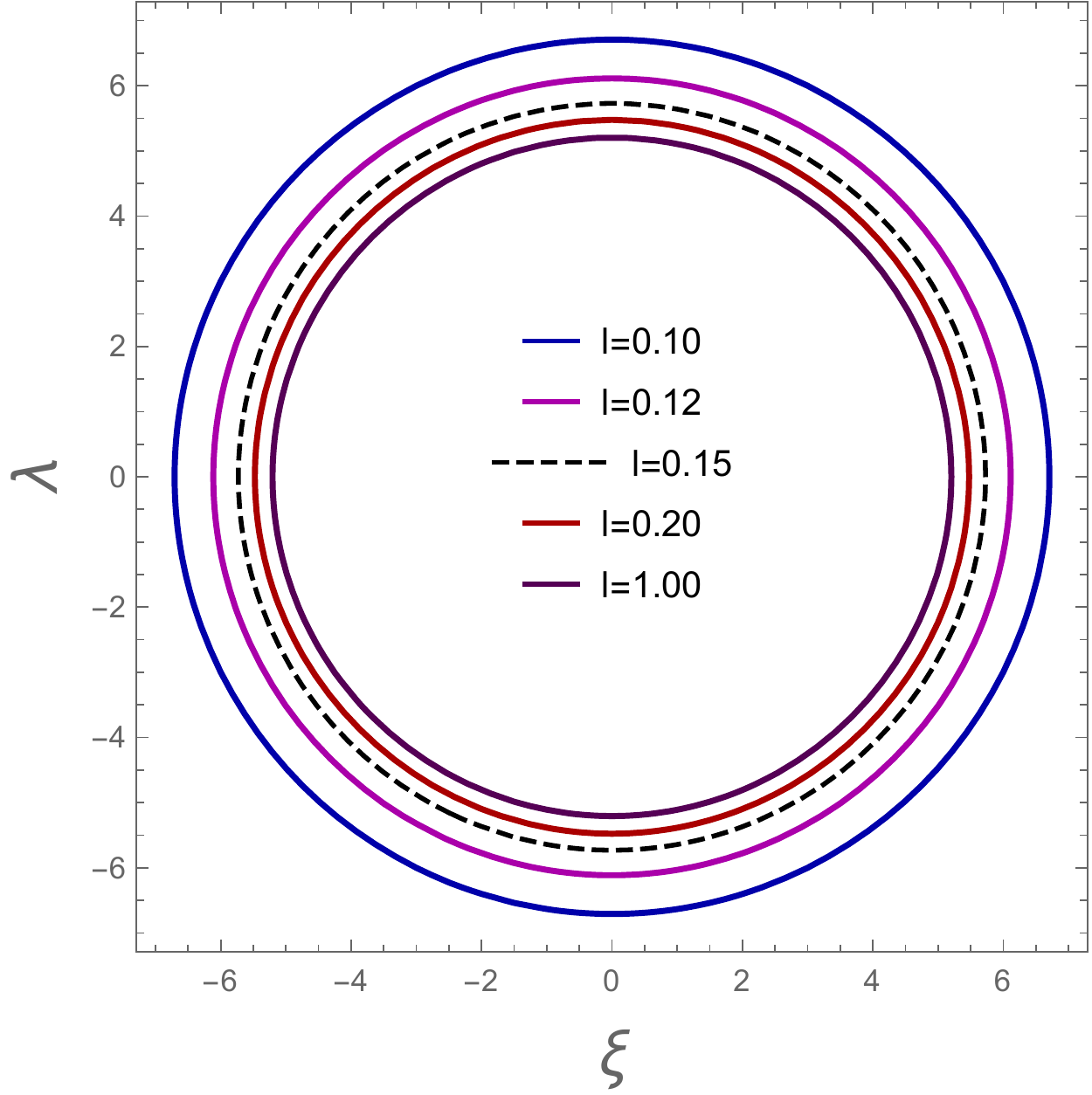}
\end{minipage}
\vspace{0.0cm}
\begin{minipage}[b]{0.58\textwidth} \hspace{-1.5cm}
\includegraphics[width=0.8\textwidth]{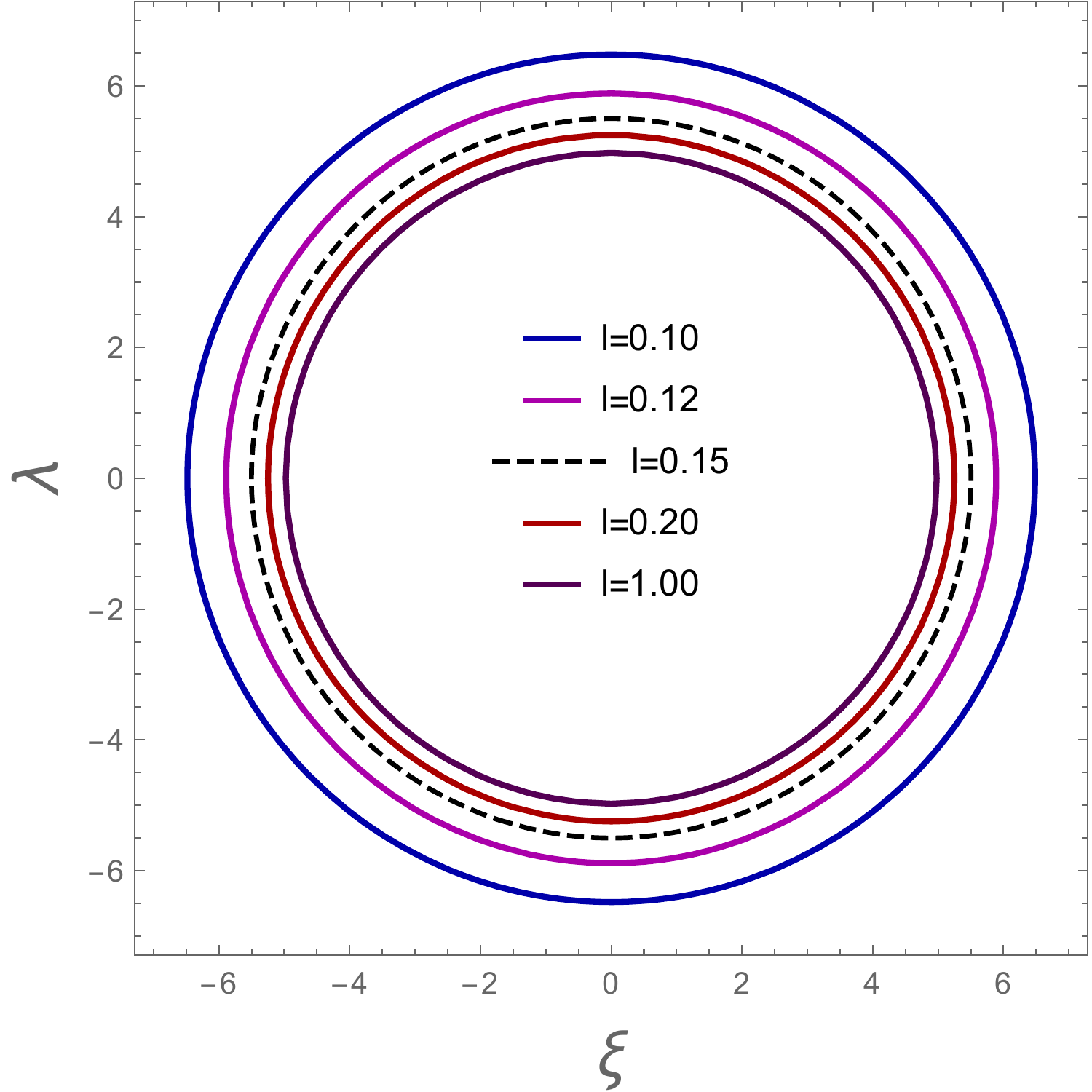}
\end{minipage}
\caption{Graphical description of BH shadow in the upper row at $l=0.8$, for $Q=0.5$ (upper left) and $\alpha=0.01$ (upper right) whereas, the lower row is plotted for $\alpha=0.001$, at $Q=0$ (left) and $Q=0.5$ (right) in case of AdS spacetime.}\label{Sh1}
\end{figure*}
\begin{equation}\label{e29}
\frac{d\phi}{dr}=\frac{L \csc^{2}\theta}{r^{2}\sqrt{\mathcal{E}^{2}-\frac{\chi(r)}{r^{2}}\left(\kappa+L^{2}\right)}},
\end{equation}
\begin{equation}\label{e30}
\frac{d\theta}{dr}=\frac{1}{r^{2}}\sqrt{\frac{\kappa - L^{2}\cot^{2}\theta}{\mathcal{E}^{2}-\frac{\chi(r)}{r^{2}}\left(\kappa+L^{2}\right)}}.
\end{equation}
Using Eqs. \eqref{e29} and \eqref{e30} at $r\rightarrow\infty$ the expression of celestial coordinates \eqref{e28}, takes the form
\begin{equation}\label{e31}
\lambda=-\frac{\gamma \csc\theta}{\sqrt{1-\frac{(\zeta+\gamma^{2})(1-\sqrt{1-\frac{4\alpha}{l^{2}}}}{2\alpha}}},
\end{equation}
\begin{equation}\label{e32}
\xi=\pm\sqrt{\frac{\zeta - \gamma^{2}\cot^{2}\theta}{1-\frac{(\zeta+\gamma^{2})(1-\sqrt{1-\frac{4\alpha}{l^{2}}}}{2\alpha}}}.
\end{equation}
For $\theta={\pi}/{2}$, the above equation simplifies to
\begin{equation}\label{e33}
\lambda=-\frac{\gamma}{\sqrt{1-\frac{(\zeta+\gamma^{2})(1-\sqrt{1-\frac{4\alpha}{l^{2}}}}{2\alpha}}},
\end{equation}
\begin{equation}\label{e34}
\xi=\pm\sqrt{\frac{\zeta}{1-\frac{(\zeta+\gamma^{2})(1-\sqrt{1-\frac{4\alpha}{l^{2}}}}{2\alpha}}}.
\end{equation}
The equation for the circle of radius in the celestial plane ($\lambda, \xi$), can be obtained by combining Eqs. \eqref{e33} and \eqref{e34}, as follows
\begin{equation}\label{e35}
R^{2}_{sh}=\lambda^{2}+\xi^{2}=\frac{\gamma^{2}+\zeta}{1-\frac{(\zeta+\gamma^{2})(1-\sqrt{1-\frac{4\alpha}{l^{2}}}}{2\alpha}}.
\end{equation} 
\begin{figure*}
\begin{minipage}[b]{0.58\textwidth} \hspace{-0.0cm}
\includegraphics[width=0.8\textwidth]{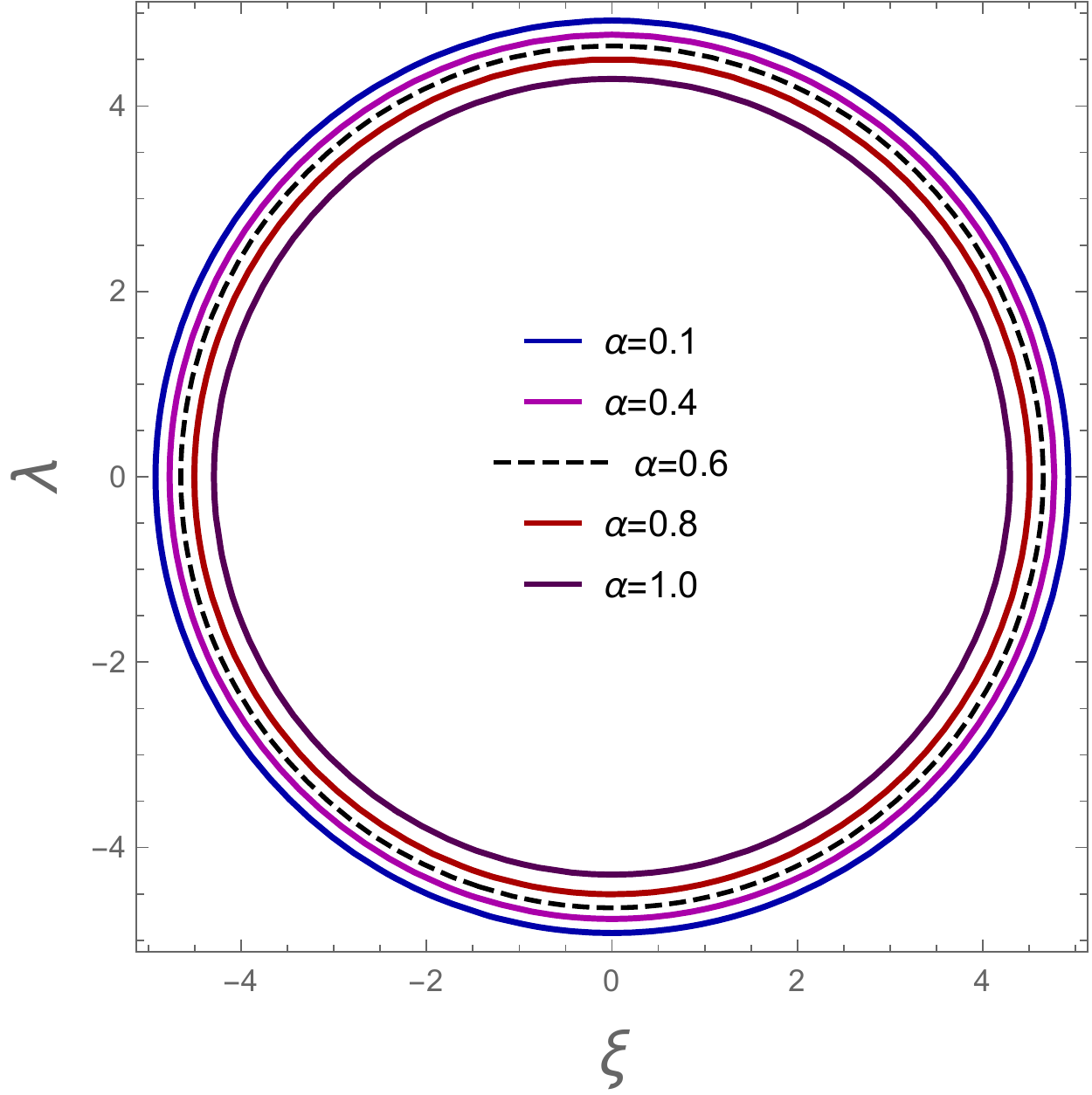}
\end{minipage}
\vspace{0cm}
\begin{minipage}[b]{0.58\textwidth} \hspace{-1.5cm}
\includegraphics[width=0.8\textwidth]{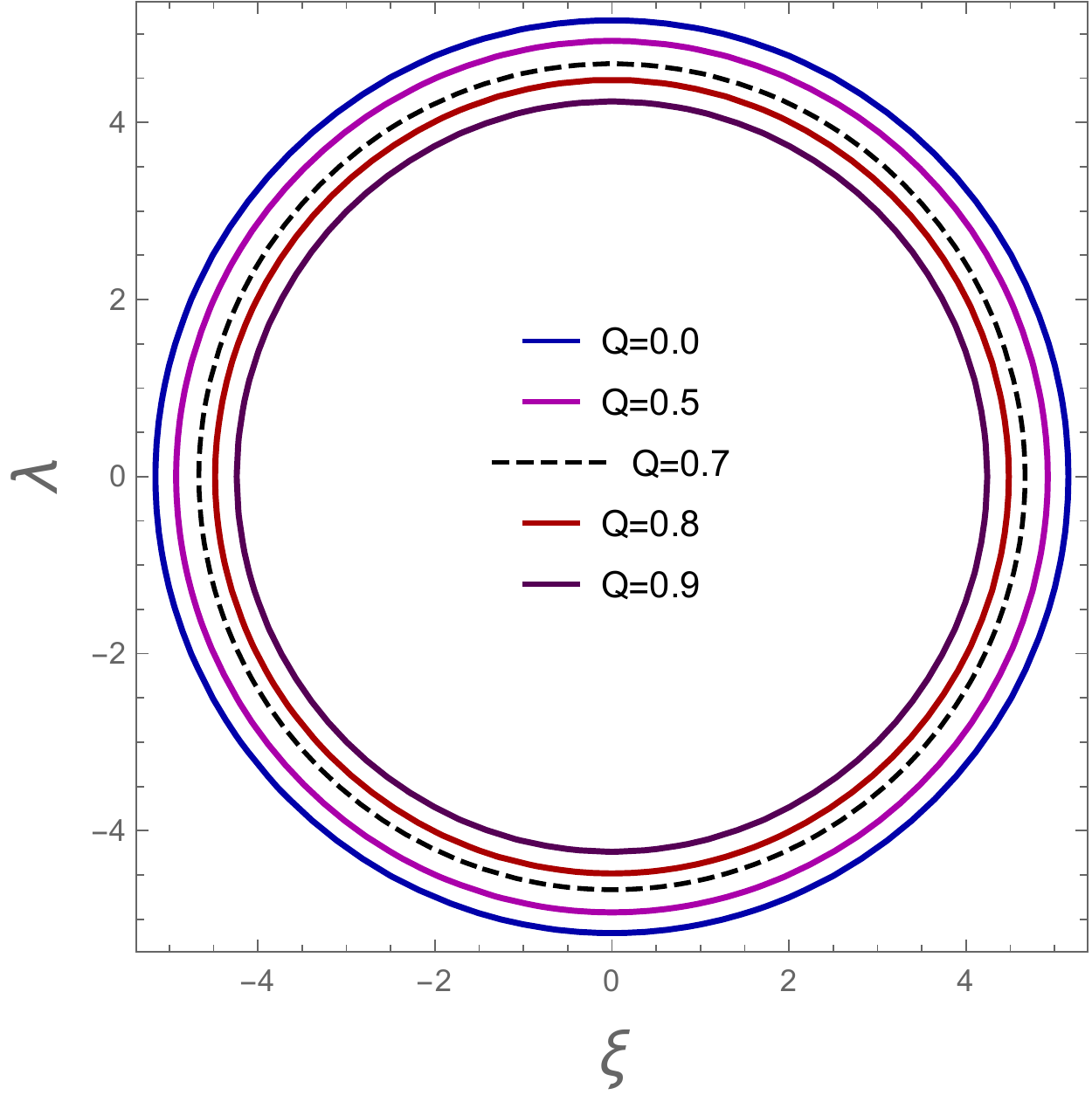}
\end{minipage}
\caption{Graphical description of BH shadow at $Q=0.5$ (left), while at $\alpha=0.1$ (right) in case of asymptotically flat spacetime BH.}\label{Sh2}
\end{figure*}
In Eq. \eqref{e35}, $R_{sh}$ represents radius of the BH shadow. Using Eq. \eqref{e22}, the radius of BH shadow can be written in the following equivalent form as
\begin{equation}\label{e36}
R_{sh}=\sqrt{\frac{\frac{r^{2}_{ph}}{\chi(r_{ph})}}{1-\frac{\left(\frac{r^{2}_{ph}}{\chi(r_{ph})}\right)\left(1-\sqrt{1-\frac{4\alpha}{l^{2}}}\right)}{2\alpha}}}.
\end{equation}
In Fig. \ref{Sh1}, we graphically illustrate the shadow cast by the GB-AdS BH under the influence of both coupling and charge parameters. From Fig. \ref{Sh1}, we observe that by increasing the value of coupling parameter causes an enlargement in BH shadow. In addition, we have also studied the effect of $l$ on the shadow radius of the BH with and without charge. From where one can clearly observe that AdS radius results in diminishing the shadow radius. Figure \ref{Sh2}, demonstrate shadow cast in case of asymptotically flat spacetime metric, where we get a contrasting result in comparison to the AdS spacetime for coupling parameter. Although, the charge parameter plays the same role in both cases and results in decreasing the shadow radius. 
\subsection{Energy emission rate}\label{sec:3.1}
In this section, we are going to find out the energy emission rate of GB-AdS BH spacetime. The general form of energy emission rate can be written as \cite{a43}
\begin{figure*}\vspace{-0.cm}
\begin{minipage}[b]{0.58\textwidth} \hspace{-0.0cm}
\includegraphics[width=0.8\textwidth]{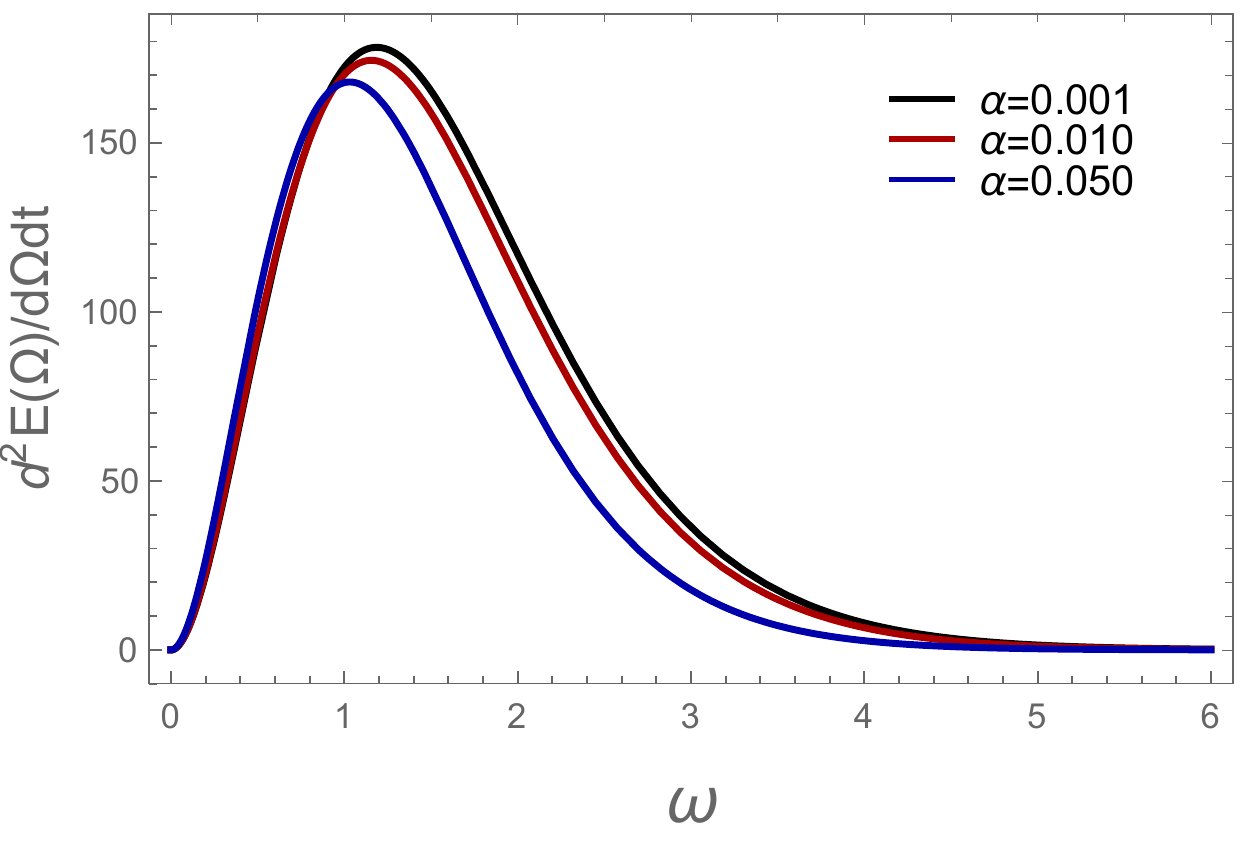}
\end{minipage}
\vspace{0.3cm}
\begin{minipage}[b]{0.58\textwidth} \hspace{-1.5cm}
\includegraphics[width=0.8\textwidth]{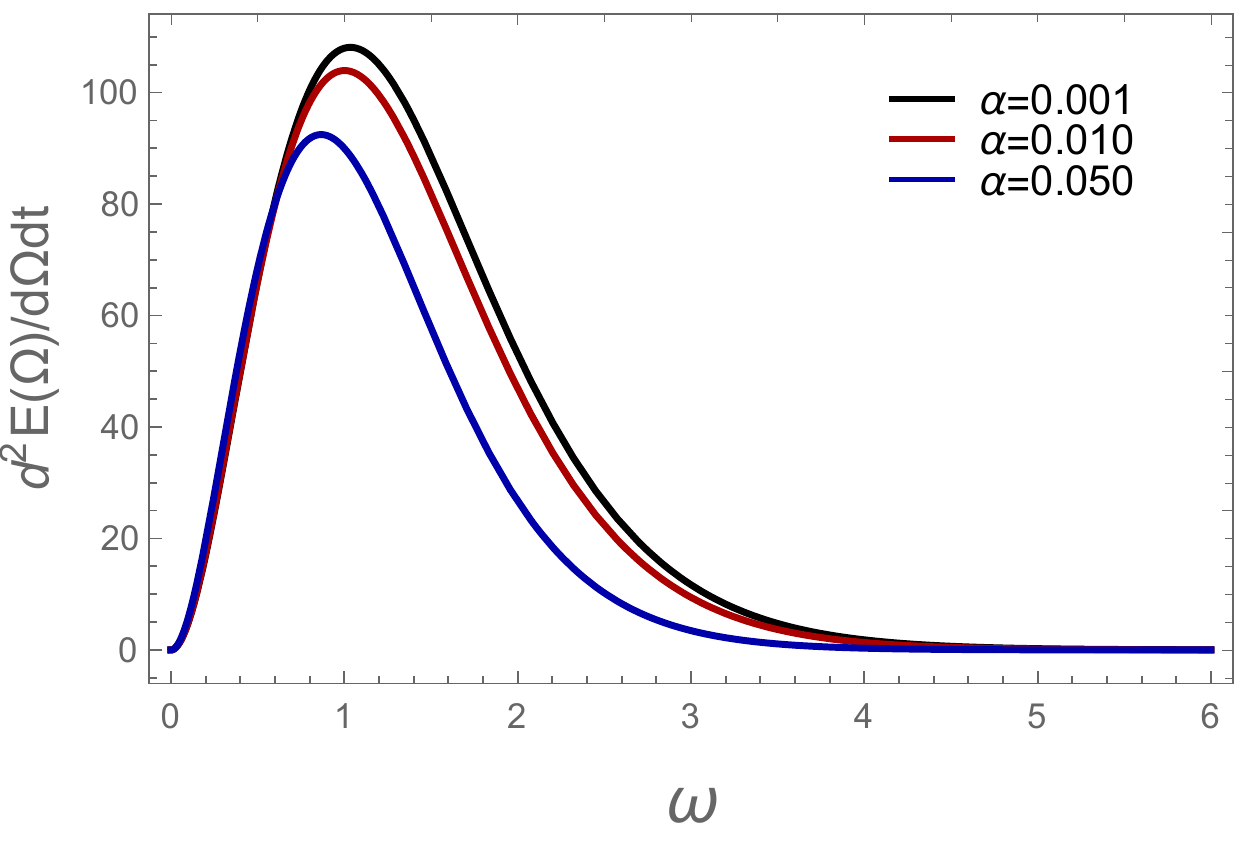}
\end{minipage}
\begin{minipage}[b]{0.58\textwidth} \hspace{3.9cm}
\includegraphics[width=0.8\textwidth]{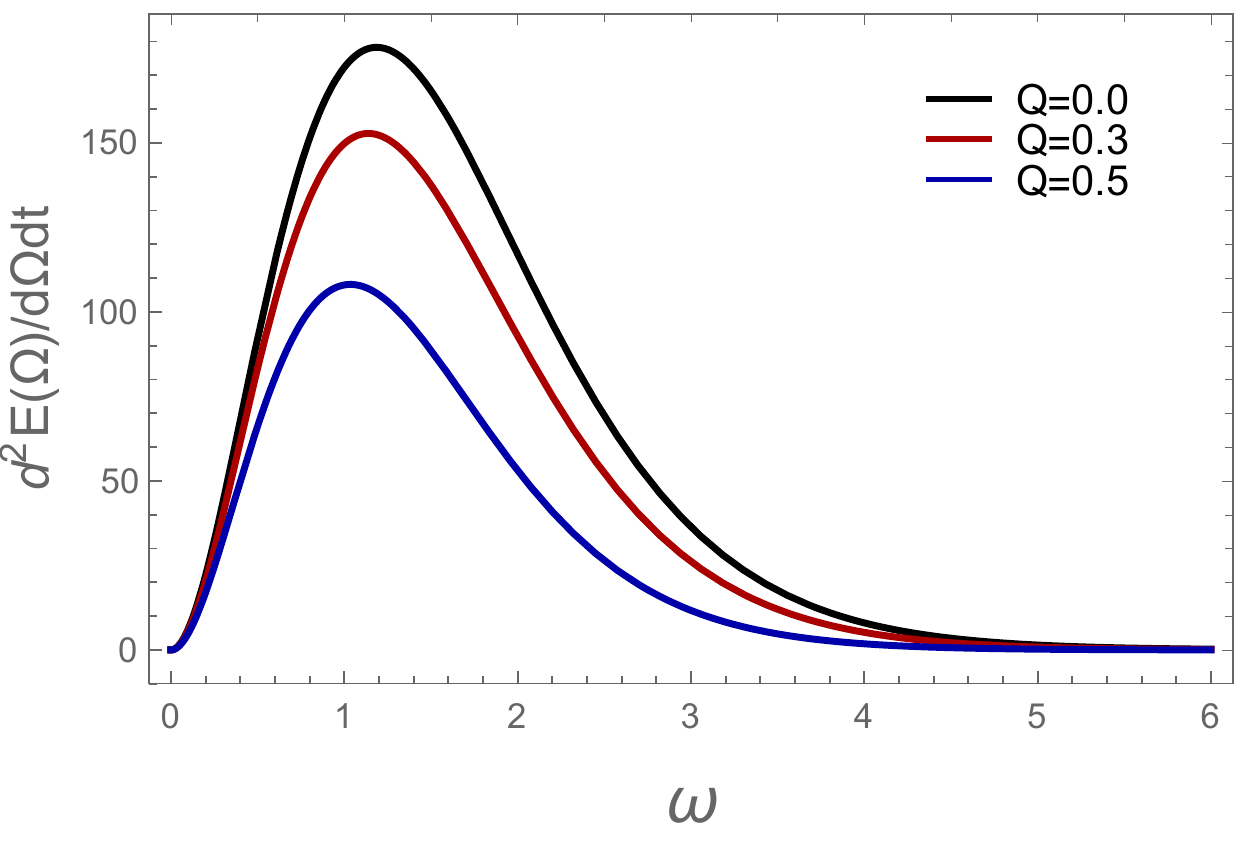}
\end{minipage}
\caption{The variation of energy emission rate vs frequency at $Q=0$ (upper left) and $Q=0.5$ (upper right), while at $\alpha=0.001$ (lower row) in case of GB-AdS spacetime ($l=0.8$).}\label{Emission1}
\end{figure*}
\begin{equation}\label{e37}
\frac{d^{2}E(\Omega)}{d\Omega dt}=\frac{2\pi^{2} \rho_{lim}}{e^{\frac{\Omega}{T_{h}}}-1}\Omega^{3}.
\end{equation}
Here $T_{h}$ , $\Omega$ and $E(\Omega)$ denotes Hawking temperature, frequency and energy of the BH, respectively. The expression for the Hawking temperature can be written as
\begin{eqnarray}\nonumber\label{e38}
T_{h}&=&\frac{\chi^{'}(r=r_{e})}{4\pi}
=\frac{4r^{3}_{e}\alpha+l^{2}\left(-2M\alpha+r^{3}_{e}\left(-1+\sqrt{1-\frac{4\alpha}{l^{2}}-\frac{4(Q^{2}-2Mr_{e})\alpha}{r^{4}_{e}}}\right)\right)}{4l^{2}\pi r^{2}_{e}\alpha\sqrt{1-\frac{4\alpha}{l^{2}}-\frac{4(Q^{2}-2Mr_{e})\alpha}{r^{4}_{e}}}}.
\end{eqnarray}
Here $r=r_{e}$, denotes the radius of the event horizon. In Eq. \eqref{e37}, $\rho_{lim}$ denotes the limiting constant value and can be expressed for the d-dimensional case as
\begin{equation}\label{e39}
\rho_{lim}=\frac{\pi^{\left(\frac{d-2}{2}\right)}R^{d-2}_{sh}}{\Gamma(\frac{d}{2})}.
\end{equation}
Therefore, for $d=4$  Eq. \eqref{e39}, becomes
\begin{equation}\label{e40}
\rho_{lim}=\pi R^{2}_{sh}.
\end{equation}
Using Eq. \eqref{e40}, in Eq. \eqref{e37}, we can get the expression for the energy emission rate as
\begin{equation}\label{e41}
\frac{d^{2}E(\Omega)}{d\Omega dt}=\frac{2\pi^{3} R^{2}_{sh}}{e^{\left(\frac{\Omega}{T_{h}}\right)}-1}\Omega^{3}.
\end{equation}
\begin{figure*}\vspace{0.0cm}
\begin{minipage}[b]{0.58\textwidth} \hspace{-0.cm}
\includegraphics[width=0.8\textwidth]{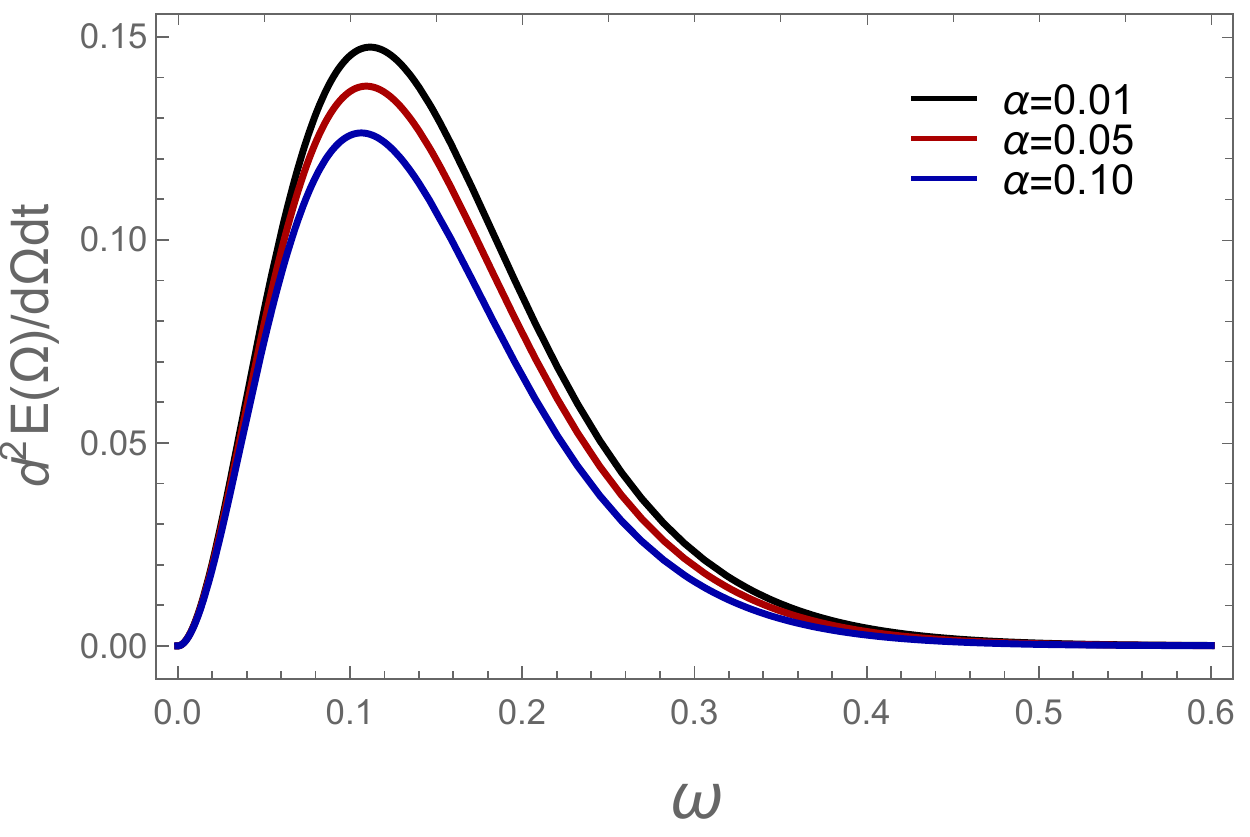}
\end{minipage}
\vspace{0cm}
\begin{minipage}[b]{0.58\textwidth} \hspace{-1.5cm}
\includegraphics[width=0.8\textwidth]{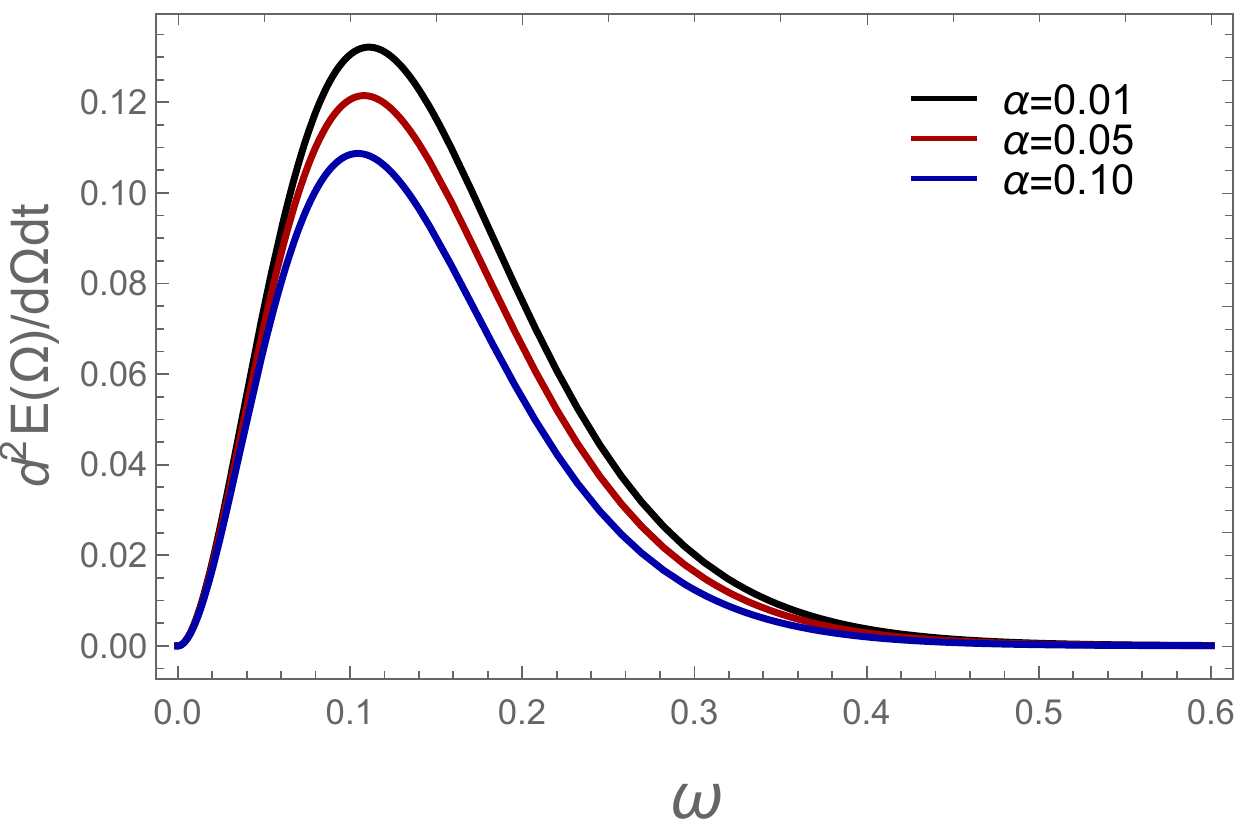}
\end{minipage}
\caption{The variation of energy emission rate vs frequency at $Q=0$ (left) and $Q=0.5$ (right) in case of asymptotically flat spacetime.}\label{Emission2}
\end{figure*}
The variations of energy emission rate along with $\Omega$, under the influence of charge and GB coupling parameter, is shown in Figs. \ref{Emission1} and \ref{Emission2}. We have checked the effect of "$\alpha$" on the energy emission rate at $Q=0$ and $Q=0.5$ (with and without charge) both for GB-AdS and asymptotically flat spacetime BHs. The plot shows that the rate of energy emission decreases by increasing the value of $\alpha$ in both charge and chargeless cases. We also observed decreasing behavior of the energy emission rate in case of charge $Q$ for the GB-AdS BH. Moreover, both AdS and asymptotically flat BHs shows same behavior of the energy emission rate, but the energy emission is smaller in asymptotically flat BH in comparison to the GB-AdS BH.
\section{Center of mass energy}
\label{sec:4}
\begin{figure*}\vspace{-0.0cm}
\begin{minipage}[b]{0.58\textwidth} \hspace{-0.0cm}
\includegraphics[width=0.8\textwidth]{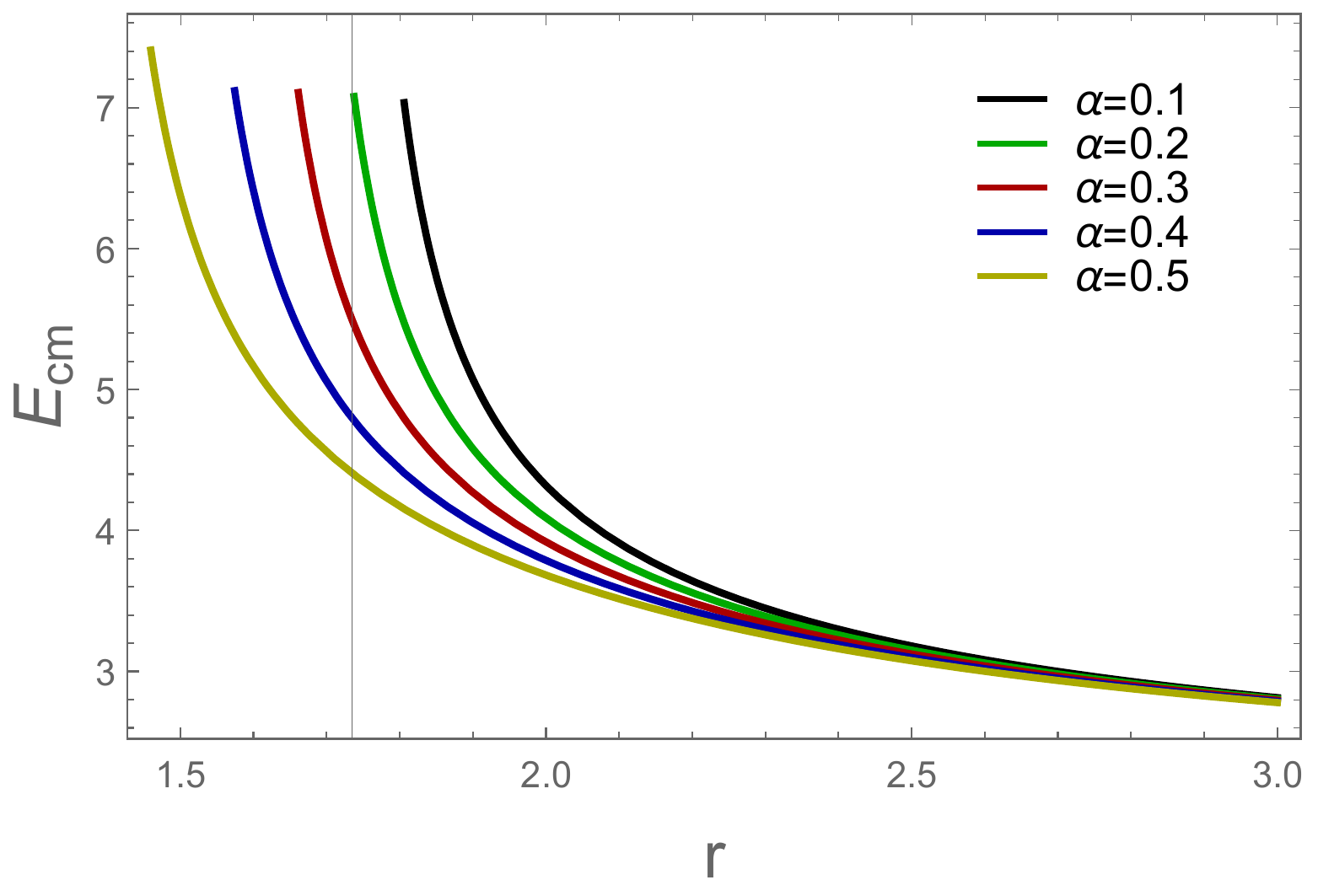}
\end{minipage}
\vspace{0.0cm}
\begin{minipage}[b]{0.58\textwidth} \hspace{-1.5cm}
\includegraphics[width=0.8\textwidth]{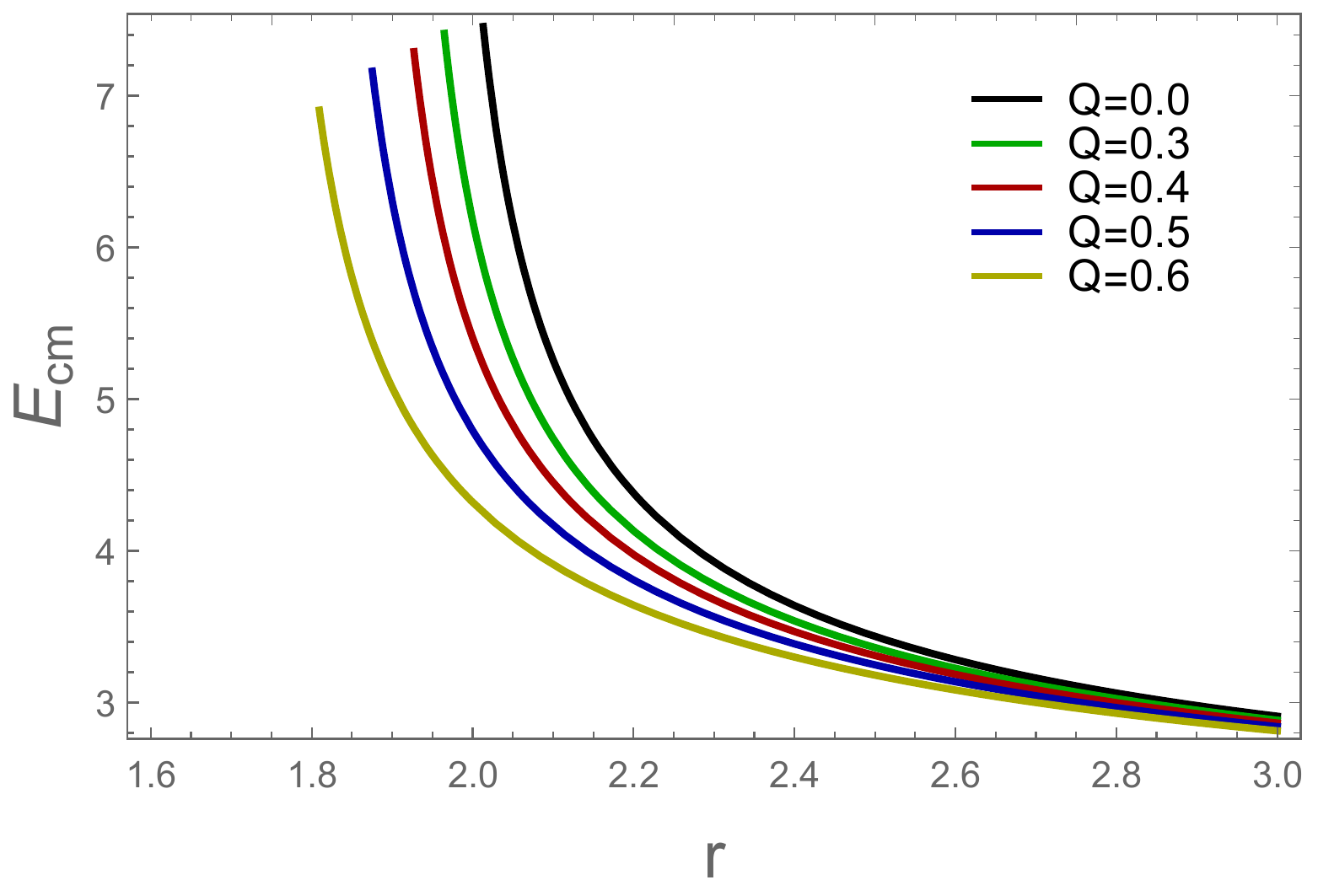}
\end{minipage}
\caption{Redial dependence of the CME for $\mathcal{E}_{1}=\mathcal{E}_{2}=1$ and $L_{1}=-L_{2}=2$,
at $Q=0.6$ (left)  and $\alpha=0.1$ (right) in case of asymptotically flat spacetime.}\label{CMenergy2}
\end{figure*}

The CME is the form of energy that is essential for the formation of new particles, as a result of colliding particles in the center of mass frame. Such energy can be achieved by adding the masses and kinetic energies of the colliding particles. Since BH can act as a particle accelerator and hence can accelerate the collision of particles up to an infinite CME. Henceforth, in this section, we are going to figure out the collision of two particles in the background of non-rotating GB-AdS BH, on an equatorial plane. For this, first of all, we need to define the energy of two particles colliding in the center of mass of the system having energies $\mathcal{E}_{1}$ and $\mathcal{E}_{2}$ at infinity. Let us assume that the two particles with equal mass ($m_{1}=m_{2}=m_{0}$) having energies $\mathcal{E}_{1}=\mathcal{E}_{2}=1$ at infinity. Now, if we denote the CME by $E_{CM}$, then by following \cite{a29}
\begin{equation}\label{e42}
E_{CM}=\sqrt{2}m_{0}\sqrt{1-g_{\mu\nu}u_{1}^{\mu}u_{2}^{\nu}}.
\end{equation}
It can be rewritten as
\begin{equation}\label{e43}
E_{CM}^{2}=2m_{0}^{2}(1-g_{\mu\nu}u_{1}^{\mu}u_{2}^{\nu}).
\end{equation}
In which $u_{i}^{\mu}$ denotes four velocities of the colliding particles for $i=1,2$. Using the values of Eqs. \eqref{e10}, \eqref{e15} and \eqref{e16}, the expression of CME takes the form
\begin{equation}\label{e44}
E_{CM}^{2}=\frac{2m_{0}^{2}}{r^{2}\chi(r)}\left(r^{2}\chi(r)+r^{2}\mathcal{E}_{1}\mathcal{E}_{2}-\chi(r)L_{1}L_{2}-P_{1}P_{2}\right),
\end{equation}
where for $i=1,2$
\begin{equation}\label{e45}
P_{i}=\sqrt{\mathcal{E}_{i}^{2}-\chi(r)\left(1+\frac{L_{i}^{2}}{r^{2}}\right)}.
\end{equation}
\begin{figure*}\vspace{0cm}
\begin{minipage}[b]{0.58\textwidth} \hspace{-0.cm}
\includegraphics[width=0.8\textwidth]{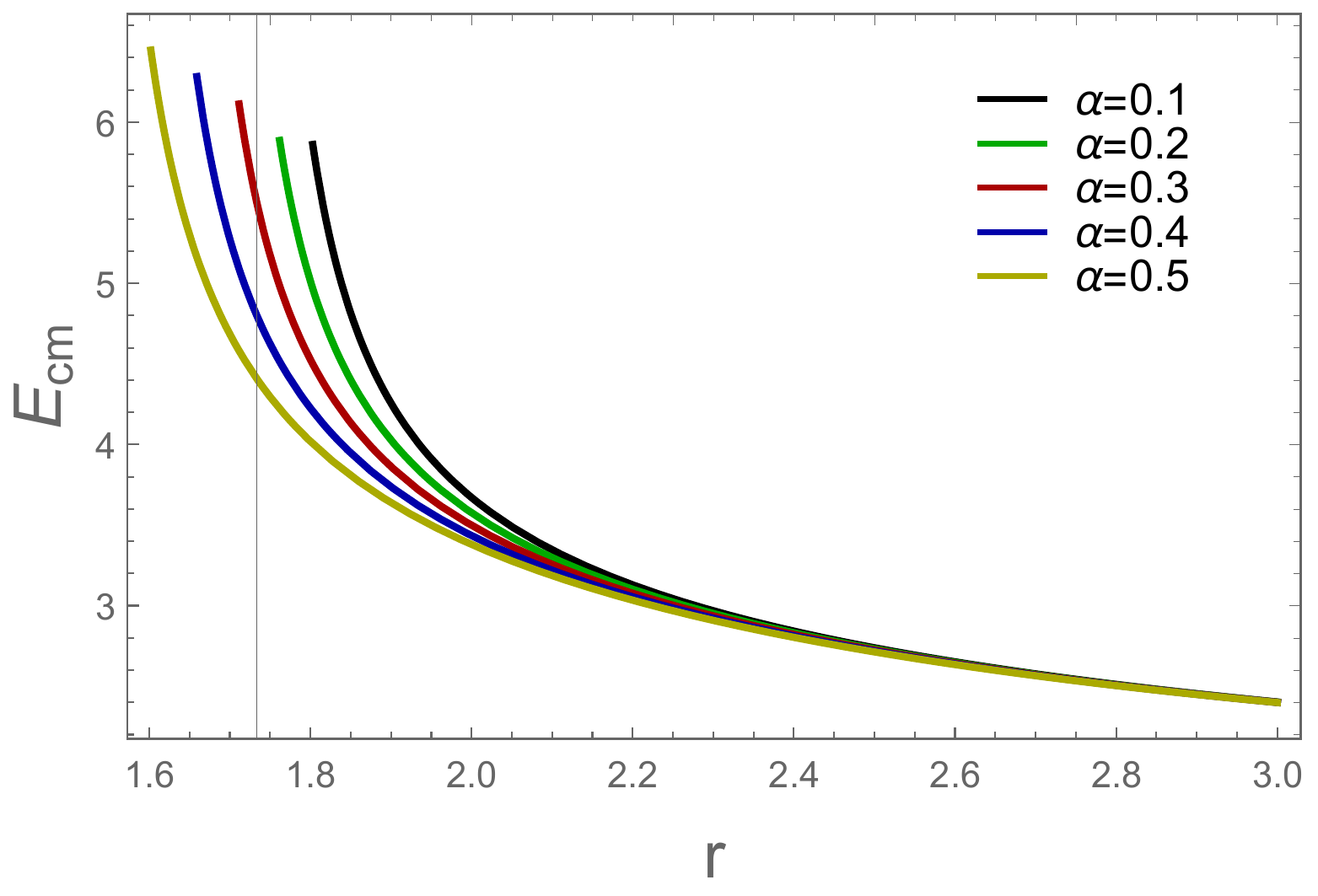}
\end{minipage}
\vspace{0.3cm}
\begin{minipage}[b]{0.58\textwidth} \hspace{-1.5cm}
\includegraphics[width=0.8\textwidth]{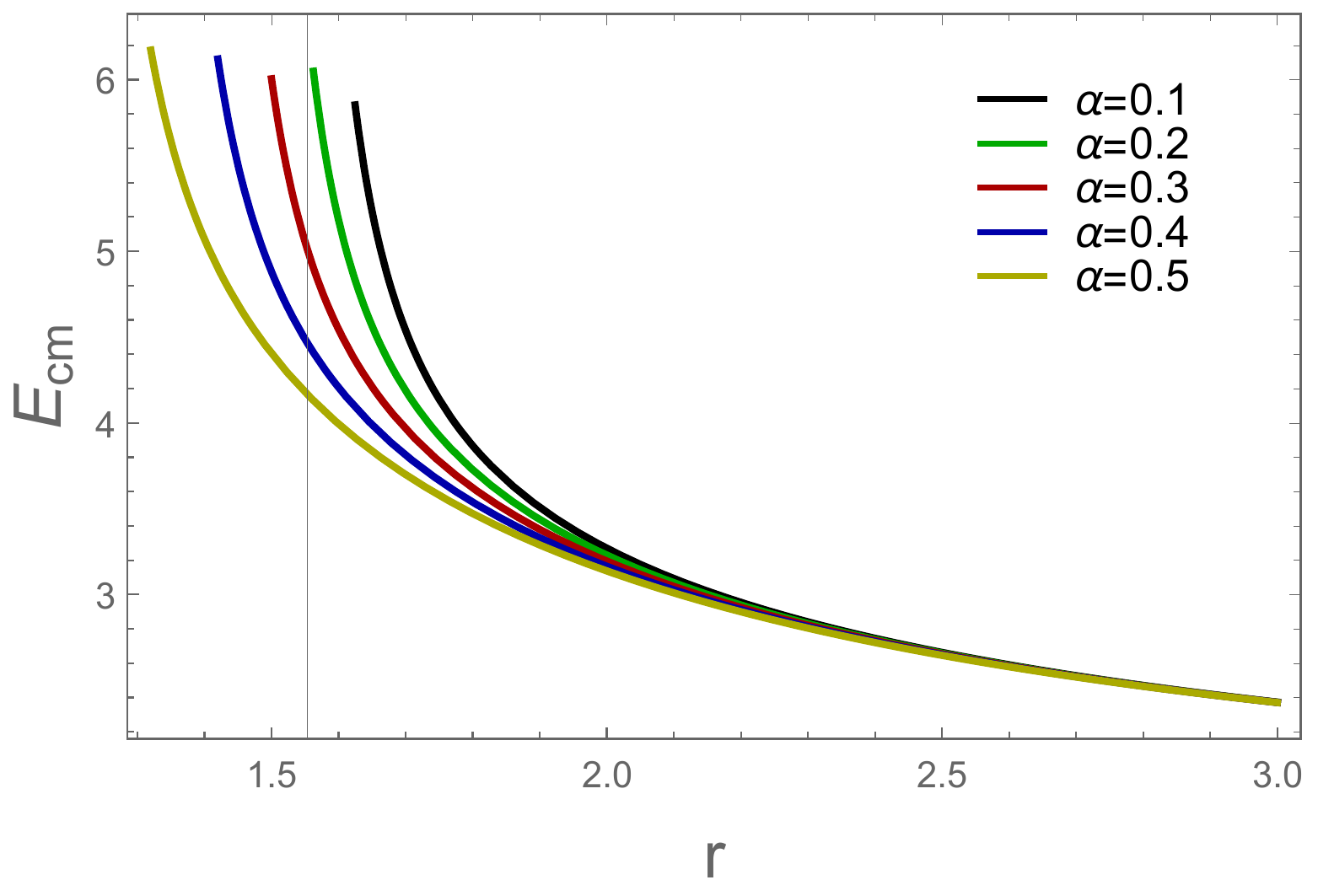}
\end{minipage}
\begin{minipage}[b]{0.58\textwidth} \hspace{-0.cm}
\includegraphics[width=0.8\textwidth]{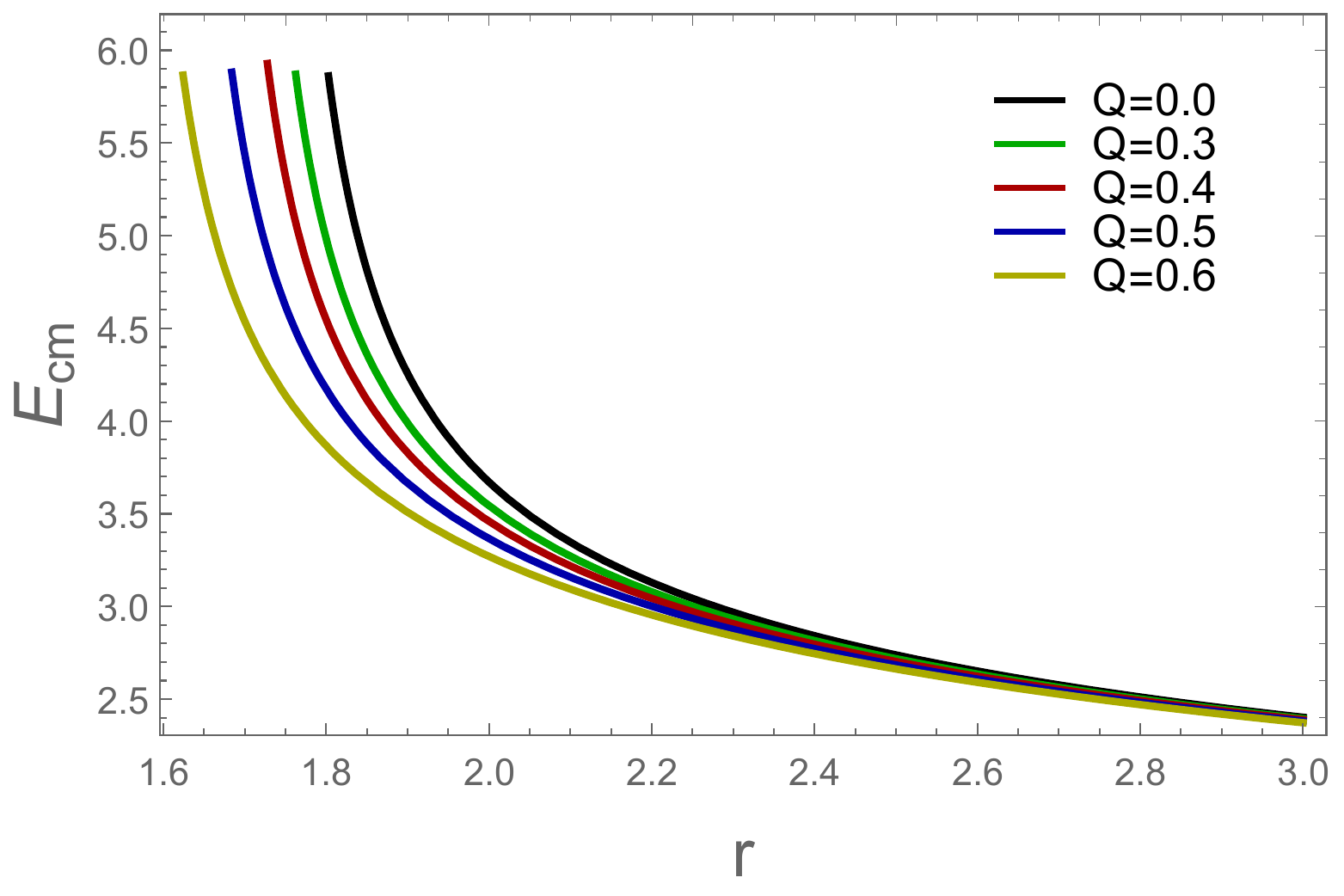}
\end{minipage}
\begin{minipage}[b]{0.58\textwidth} \hspace{-1.5cm}
\includegraphics[width=0.8\textwidth]{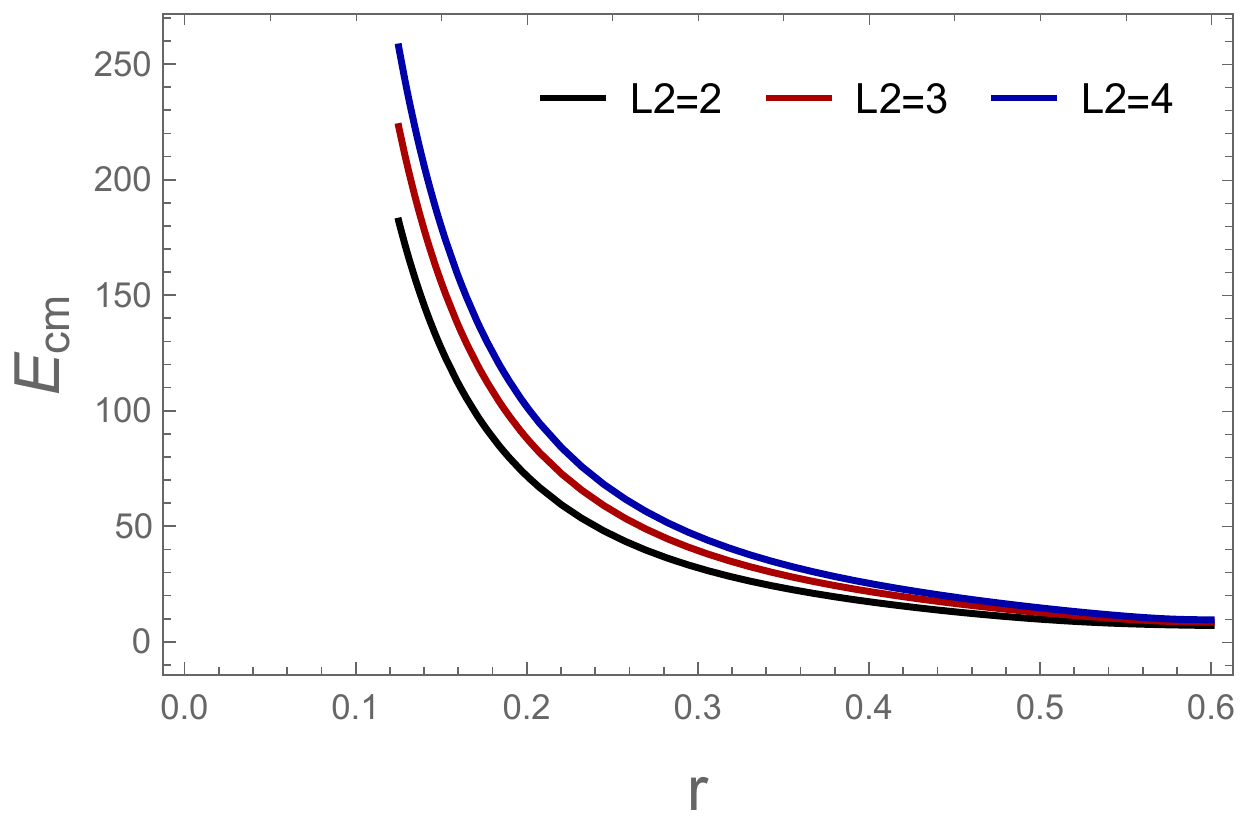}
\end{minipage}
\caption{Redial dependence of the CME for $\mathcal{E}_{1}=\mathcal{E}_{2}=1$ and $L_{1}=-L_{2}=2$,
at $Q=0$ and $Q=0.6$ (upper left and right panels) whereas, at $\alpha=0.1$ (lower left panel) in case of GB-AdS BH.}\label{CMenergy1}
\end{figure*}
The radial dependence of CME at different values of $\alpha$ and $Q$ is shown in Figs. \ref{CMenergy1} and \ref{CMenergy2}. The Fig. \ref{CMenergy1} is plotted at AdS radius $l=5$ and the variation of other parameters. From the graphical representation, one can clearly observe that CME diminishing with both charge and coupling parameters. Additionally, $L_{2}$ result in increasing the value of CME. In case of the asymptotically flat spacetime BH, the graphical behavior of CME shows the same behavior as that of GB-AdS BH but possess greater value of the CME.
\section{Conclusion}
\label{sec:5}
The purpose of this work is to investigate the shadow and CME of a 4-dimensional charged GB-AdS spacetime BH. We expect that on the observational point of view, our exploration will be much beneficent for future EHT observations. For this purpose, first, we construct the geodesic equations of a particle moving in the background of GB-AdS spacetime. We then studied the horizon structure of BH and explored its response against the charge and GB coupling parameter. We found that both charge and coupling parameters result in denser the BH, as it diminishes its horizons.

To find the radius of BH shadow, we first derived the expressions of celestial coordinates using null geodesics. We then graphically construct BH's shadow for both of AdS and asymptotically flat spacetime BHs at different values of the BH parameters. Our finding shows that both charge, as well as the GB parameter, affects the shadow of BH. The acquired result shows that in the case of AdS spacetime, $l$ decreases the shadow radius, whereas $\alpha$ results in an increase of the shadow radius. On the other hand, $\alpha$ results in increases the shadow radius in asymptotically flat spacetime. Besides, the charge parameter leads to shrank BH's shadow in both GB-AdS and asymptotically flat spacetimes.

After that, we studied and graphically depicted the rate of energy emission in the vicinity of GB-AdS spacetime. The results indicate that upon increasing the values of both $Q$ and $\alpha$, the energy emission rate gets diminished. Similar behavior is observed in the case of asymptotically flat spacetime as well. 
Since it is a well-known fact that a BH can act like a particle accelerator. In this work, we have computed the CME on the equatorial plane using particle's equations of motion. For this purpose, we consider the collision of two particles in the locality of GB-AdS BH. The obtained results show that CME increases by increasing the GB parameter. While a similar behavior can also be observed by varying the value of BH charge.     
\subsubsection*{Declaration of competing interest}
The authors declare that they have no known competing financial interests or personal relationships that could have appeared
to influence the work reported in this paper.
\subsubsection*{Acknowledgment}
This work is supported by the NSFC Project (11771407) and the MOST Innovation Method Project (2019IM050400).

\end{document}